\documentclass[12pt, letterpaper]{article}

\usepackage[utf8]{inputenc}
\usepackage[T1]{fontenc}
\usepackage[english]{babel}
\DeclareUnicodeCharacter{2191}{\ensuremath{\uparrow}}
\usepackage{mathptmx}
\usepackage[margin=1in]{geometry}
\usepackage{setspace}
\usepackage{amsmath}

\usepackage{amssymb}
\usepackage{graphicx}
\usepackage{xcolor}
\graphicspath{{./}{paper_outputs/}}
\usepackage[numbers]{natbib}
\usepackage{booktabs}
\usepackage{longtable}
\usepackage{microtype}
\usepackage{caption}
\usepackage{hyperref}
\usepackage{url}
\definecolor{arxivblue}{rgb}{0.0, 0.2, 0.65}
\hypersetup{
	pdftitle={Bitcoin's Structural Position as Money: A Contested Synthesis, Revised},
	pdfauthor={Hamoon Soleimani},
	pdfkeywords={Bitcoin, Monetary Theory, Post-Keynesian Economics, Austrian School, Regression Theorem, Volatility, Lightning Network, 51 percent attack, Basel III},
	colorlinks=true, linkcolor=arxivblue, citecolor=arxivblue, urlcolor=arxivblue,
	breaklinks=true, pdfstartview={FitH}, pdfcreator={pdflatex}
}

\title{Bitcoin's Structural Position as Money: A Contested Synthesis of Post-Keynesian and Austrian Critiques}

\author{Hamoon Soleimani}
\date{September 5, 2026}

\begin{document}
	\maketitle
	\thispagestyle{empty}
	
\begin{abstract}
	Since its inception, Bitcoin has been positioned as a revolutionary alternative to national
	currencies. This paper evaluates that claim against two competing theoretical traditions --
	Post-Keynesian monetary theory and the Austrian School -- while evaluating the strongest 
	counter-arguments within each tradition rather than treating either verdict as settled. 
	From a Post-Keynesian perspective, Bitcoin lacks the debt-based IOU architecture and 
	state-enforced fiscal acceptance mechanism that anchors sovereign currency \citep{vianna2021}, 
	though this framework faces definitional circularity when applied to non-debt commodity monies 
	such as gold. From an Austrian viewpoint, Bitcoin's consistency with Mises's Regression Theorem 
	remains deeply contested: while early-adopter subjective utility can theoretically resolve the 
	regression problem \citep{davidsonblock2015, peniaz2024}, a strict reading demonstrates that it 
	lacks an independent, non-monetary commodity anchor \citep{hazlettluther2020, umlauft2018}. 
	
	These theoretical frameworks are tested against empirical evidence across settlement throughput, 
	Layer-2 Lightning Network routing dynamics, 51\% attack economics, market microstructure, and 
	sovereign adoption in El Salvador. We demonstrate that while Lightning Network routing reliability 
	has expanded via Pickhardt--Richter multi-path flow allocations, this performance gain is 
	structurally concentrated in well-capitalized custodial liquidity hubs --
	confirming the game-theoretic centralization predicted by Avarikioti et al.~\citep{avarikioti2020}. The paper concludes that Bitcoin's architectural trade-offs, rather than superficial operational metrics, systematically 
	impede its capacity to function as a sovereign monetary standard.
\end{abstract}
	
	\vspace{1em}
	\noindent\textbf{Keywords:} Bitcoin, Cryptocurrencies, Monetary Theory, Post-Keynesian Economics,
	Austrian School of Economics, Regression Theorem, Volatility, Scalability, Systemic Risk, Lightning
	Network, 51\% Attacks.
	\vspace{1.5em}

	\section{Introduction: The Bitcoin System and the Central Research Question}
	Bitcoin emerged in 2009 from a white paper authored by the pseudonymous Satoshi Nakamoto, which
	introduced a "peer-to-peer electronic cash system" intended to facilitate online payments "without
	going through a financial institution" \citep{nakamoto2008}. Arriving in the aftermath of the 2008
	global financial crisis, it was quickly championed as a decentralized alternative to state-controlled
	fiat currencies.
	
	At its technical core, the protocol solves the "double-spending problem" through a public,
	decentralized ledger -- the blockchain -- maintained by a distributed network of participants,
	removing the need for a trusted intermediary \citep{nakamoto2008}. Network security is maintained
	through "proof-of-work": miners compete to solve a cryptographic puzzle, and the first to succeed
	appends a block and receives newly issued bitcoin plus transaction fees, which supplies the economic
	incentive to secure the network \citep{lo2014bitcoin}.
	
	Two design choices are central to Bitcoin's economic identity. First, its supply is capped at 21
	million units, with issuance halving roughly every four years, engineered to create scarcity
	analogous to a precious metal. Second, the system is pseudonymous and permissionless. These features
	underpin the "digital gold" and "inflation hedge" narratives -- narratives that do not hold up
	uniformly under scrutiny: wavelet-coherence analysis of Bitcoin against forward inflation
	expectations finds a positive relationship only during the acute 2020 COVID shock, not as a general
	property \citep{conlon2021}.
	
	This paper evaluates Bitcoin against two competing schools of monetary theory -- the Post-Keynesian
	framework, which treats money as a debt-based social institution, and the Austrian School, which
	treats money as an evolved market commodity -- and then tests the resulting predictions against
	current empirical data. This evaluation does not treat either school's harshest reading as an 
	uncontested conclusion: where a school is internally divided (as the Austrian School is on Bitcoin 
	and the Regression Theorem), that division is reported, and the paper's own choice of which reading 
	to proceed with is stated explicitly as a methodological assumption, open to being wrong.
	
	\subsection{Literature Review}
	Yermack \citep{yermack2014} argues Bitcoin's volatility undermines its candidacy as medium of
	exchange, unit of account, and store of value simultaneously. Lo and Wang \citep{lo2014bitcoin}
	reach a similar verdict, emphasizing that Bitcoin's value is driven more by speculation than
	fundamental monetary use. Budish \citep{budish2018} provides the formal economic backbone this paper
	leans on for its security-budget argument: proof-of-work security requires that the perpetual
	"flow" of miner revenue exceed the one-off "stock" value obtainable from a successful attack,
	$P_{block} > V_{attack}/\alpha$, which is a structural, not incidental, tension once the flow itself
	must decline (Sections \ref{sec:security} and \ref{sec:governance}).
	
	Vianna \citep{vianna2021} argues from a Post-Keynesian standpoint that Bitcoin lacks the debt-IOU
	structure of modern money. A significant counter-argument comes from Hazlett and Luther's own later
	work \citep{hazlettluther2020}, which argues the standard multi-function definition of money
	(medium of exchange \emph{and} store of value \emph{and} unit of account) is too restrictive, and
	that under a narrower medium-of-exchange definition, Bitcoin's demand-to-hold already exceeded that
	of the large majority of the world's 106 sovereign currencies as of 2018 -- a genuinely
	pro-Bitcoin finding this paper reports in full rather than omitting (Section
	\ref{sec:countervailing}).
	
On scalability, early empirical analyses relied on a 2020 active-probing study of
Lightning Network reliability \citep{waughholz2020}. That study's own numbers are accurately quoted
here, but citing them alone as the network's \emph{current} state is no longer defensible: Pickhardt and
	Richter's minimum-cost-flow routing framework \citep{pickhardtrichter2021} has since been deployed in
	production by Core Lightning (\texttt{renepay}, since v23.08) and LND (bimodal balance model, since
	v0.19), and River Financial's own platform reports a 99.7\% payment success rate across 308{,}000
	transactions as of August 2023 \citep{river2023}. Section \ref{sec:ln} explains why both facts are
	true simultaneously and why the second does not rebut this paper's centralization thesis --
	if anything it is the thesis's own prediction, arriving on schedule.
	
	\section{The Theoretical Nature of Money: A Dual-Framework Analysis, With Its Own Dissents}
	\subsection{The Post-Keynesian Framework: Money as a State-Enforced IOU}
	From a Post-Keynesian perspective, money is a social relationship of credit and debt rather than a
	commodity. Minsky's observation that "everyone can create money; the problem is to get it accepted"
	\citep{minsky1986} captures the resulting "hierarchy of money," in which the state's own IOU sits at
	the apex because the state can impose tax liabilities extinguishable only in its own currency
	\citep{wray2015}. Vianna \citep{vianna2021} extends this to argue Bitcoin fails as money because it
	is not an IOU: there is no issuer and no corresponding liability.
	\begin{figure}[hbt!]
		\centering
		\includegraphics[width=0.8\textwidth]{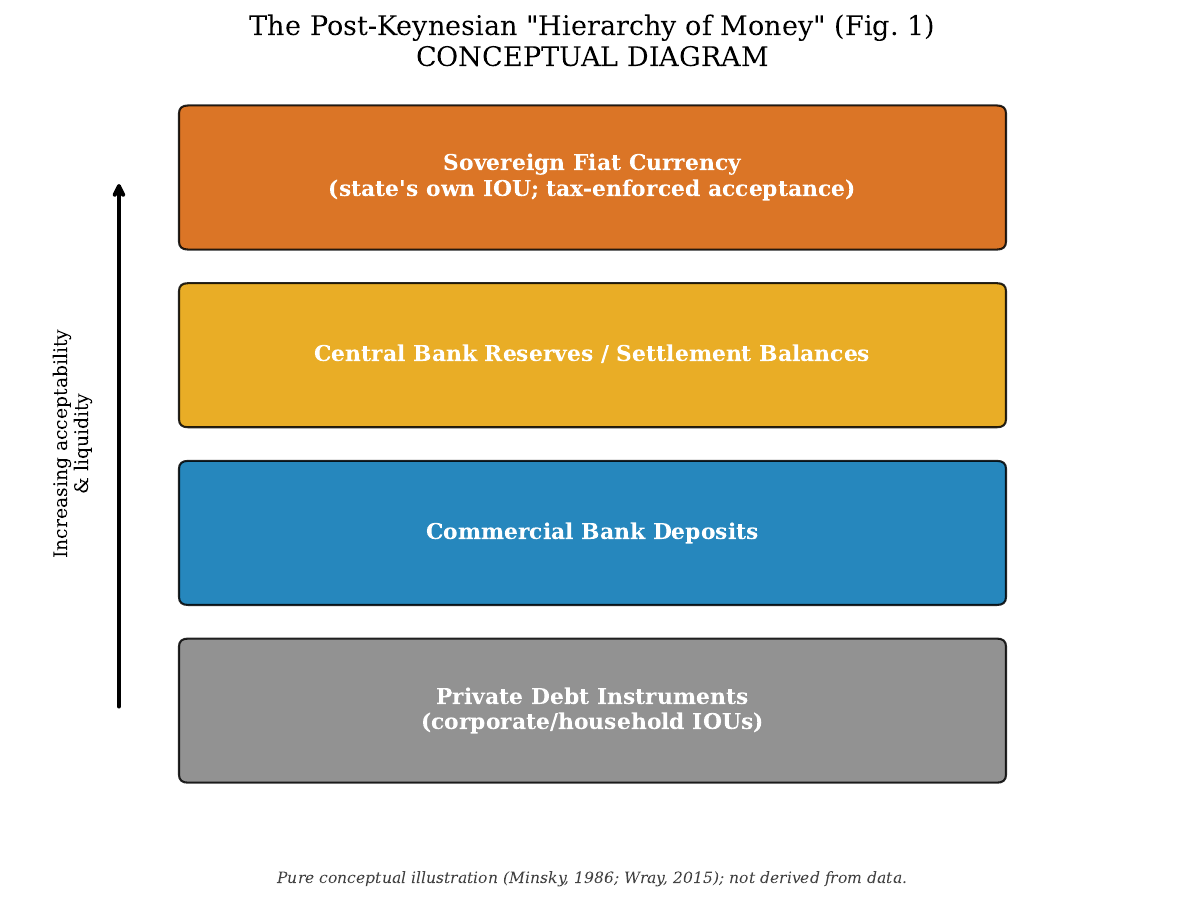}
		\caption{The Post-Keynesian hierarchy of money (conceptual diagram).}
		\label{fig:hierarchy}
	\end{figure}
	
	\emph{A necessary caveat.} Taken to its logical end, this argument proves too much: a commodity
	money with no issuer -- gold under a classical gold standard -- is not an IOU either, yet
	historically functioned as money for millennia, including as the reserve asset \emph{underlying}
	state IOUs. The Post-Keynesian response is that gold's monetary role was itself parasitic on
	state-run mints and legal-tender laws that made it the settlement medium for state-enforced debts,
	so the tax-backing argument is not strictly circular once the state's coercive power over commodity
	\emph{acceptance} (not just currency issuance) is included. We report this rebuttal for completeness
	but flag that it shifts the argument from "money must be an IOU" to the narrower and more defensible
	"money's durability as a general medium of exchange is easier to explain, and historically has been
	better explained, when a coercive institution stands behind its acceptance" -- a claim compatible
	with commodity monies but still unfavorable to an asset, like Bitcoin, with neither an issuer nor a
	coercive acceptance mechanism.
	
	\subsection{The Austrian School Framework: Money as the Most Saleable Commodity}
	Carl Menger held that money emerges spontaneously from barter as market participants gravitate
	toward whichever commodity is most saleable (\textit{Absatzf\"ahigkeit}) \citep{menger1892}. Ludwig
	von Mises's Regression Theorem addresses the resulting circularity problem -- money is valued
	because it has purchasing power, but has purchasing power because it is valued -- by requiring that
	a money's value be traceable back to a point of non-monetary use \citep{mises1953}.
	
\subsubsection{The regression-theorem dispute}
Applied to Bitcoin, this framework is internally contested. Under a strict, orthodox reading, 
Bitcoin was created \emph{ex nihilo} with no prior non-monetary use value comparable to gold's 
industrial or ornamental demand; Umlauft \citep{umlauft2018} defends this reading and additionally 
argues that Bitcoin's value is often justified through an "Input Fallacy of Value" -- the mistaken 
idea that mining's energy cost itself confers value, a modern echo of the labor theory of value 
the Austrian tradition rejects.

Against this, Davidson and Block \citep{davidsonblock2015} argue the theorem's non-monetary-use
requirement applies only to the very \emph{first} money to emerge from a pure barter economy. 
Peniaz and Kavaliou \citep{peniaz2024} independently reach a consistent conclusion: because early 
adopters subjectively valued Bitcoin and exchanged it within an existing monetary context (deriving 
direct utility from censorship resistance, cryptographic verification, and network experimentation), 
the asset does not formally violate the Regression Theorem.

Given this foundational division -- Umlauft's \citep{umlauft2018} strict commodity-anchor 
requirement versus the subjective-network-utility framework of Davidson and Block 
\citep{davidsonblock2015} and Peniaz and Kavaliou \citep{peniaz2024} -- this paper evaluates 
Bitcoin's monetary candidacy under the \emph{strict} reading as an explicit methodological 
baseline. The Regression Theorem's primary analytical function is to discipline self-referential 
asset valuation; a purely speculative, forward-looking asset represents the boundary case this 
theorem was designed to interrogate. Readers who adopt the permissive subjective-utility 
interpretation should view the Austrian critique as conditional, directing analytical focus toward 
the Post-Keynesian institutional critique and the empirical market structure sections 
(Sections 3--9).
	
	Under the strict reading, the Austrian critique is twofold: (1) inconsistency with the Regression
	Theorem as argued above, and (2) failure to emerge as Menger's "most saleable commodity" -- the
	overwhelming majority of global transactions remain priced and settled in fiat, and Austrian
	economists distinguish a limited \textit{medium of exchange} (which Bitcoin is, narrowly) from
	\textit{money proper} (the single most widely accepted medium, which Bitcoin is not)
	\citep{menger1892,mises1953}. Section \ref{sec:elsalvador} reports the one direct natural experiment
	testing whether legal-tender \emph{decree} can substitute for Mengerian market selection.
	
	\subsection{A Countervailing View: Defining Money by Exchange Functionality Alone}
	\label{sec:countervailing}
	It is important to also report the strongest available pro-Bitcoin theoretical argument in full,
	rather than only its critiques. Hazlett and Luther's later work \citep{hazlettluther2020} argues
	against the conventional multi-function definition of money used by Yermack and by this paper's own
	Section 2.2: they contend the historically and analytically prior definition of money is simply "a
	commonly accepted medium of exchange," and that volatility or store-of-value performance may affect
	an asset's \emph{quality} as money without disqualifying it from the category altogether. Under this
	narrower test, Hazlett and Luther compare Bitcoin's market capitalization against 106 sovereign base
	monies and find that, in 2018, Bitcoin's mean market capitalization of \$129.27 billion exceeded that
	of all but six national currencies -- leading them to conclude Bitcoin is "worthy of the label
	money, if only over a relatively small domain."
	
	This argument is genuinely strong and is not fully answered by pointing to volatility, since
	volatility is explicitly bracketed out of their definition. The weaker point in their argument, and
	the one this paper leans on instead, is that market capitalization conflates \emph{speculative}
	demand (a bet on future price appreciation) with \emph{transactional} demand (routine use as a
	payment medium) -- a distinction the wash-trading and stablecoin-dependency evidence in Section
	\ref{sec:wash} and the El Salvador natural experiment in Section \ref{sec:elsalvador} are intended
	to probe directly, rather than asserting by definition.
	
	\section{Econometric Analysis of Extreme Volatility}
	\label{sec:volatility}
	An asset lacking a sovereign value anchor and possessing a perfectly inelastic supply should exhibit
	extreme price instability; this section tests that prediction. Chinazzo and Jeleskovic
	\citep{chinazzo2024} compare historical, GARCH-forecasted, and options-implied volatility for Bitcoin
	and find a "notably high expected level of volatility" is a persistent structural feature.
	
	\subsection{Comparative Volatility Metrics}
	Recent estimates put Bitcoin's annualized historical volatility above 100\% during active periods, an
	order of magnitude greater than major fiat currencies or gold; even in calmer periods it runs
	roughly 3.6$\times$ gold's volatility and 5.1$\times$ global equities' \citep{baek2014,chinazzo2024}.
	
	\begin{figure}[hbt!]
		\centering
		\includegraphics[width=\textwidth]{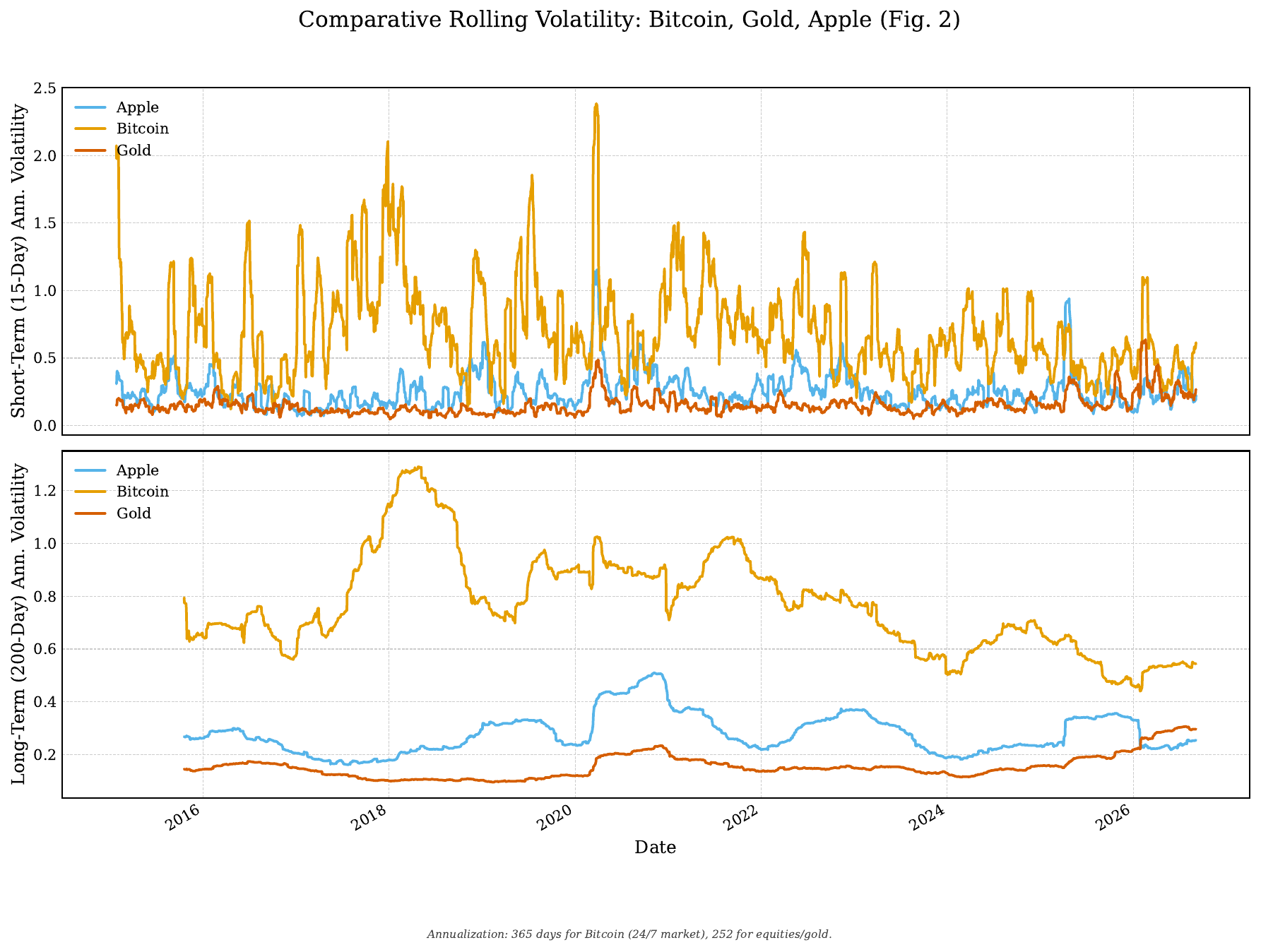}
		\caption{Short-term (15-day) and long-term (200-day) rolling volatility for Bitcoin, Gold, and
			Apple, annualized using asset-specific trading-day conventions ($N=365$ for Bitcoin, which
			trades every calendar day; $N=252$ for equities and gold).}
		\label{fig:vol_comparison}
	\end{figure}
	
	\subsection{Value-at-Risk Analysis}
Using a standard 1-day 95\% confidence interval, Value-at-Risk is reported under
three methods -- historical simulation, a Cornish-Fisher parametric adjustment for skewness and	kurtosis, and a Peaks-Over-Threshold Extreme Value Theory (EVT) fit of a Generalized Pareto
	Distribution \citep{mcneilfrey2000} -- each paired with Expected Shortfall, and each historical-VaR
	model is backtested out-of-sample with a rolling 250-day window using the Kupiec \citep{kupiec1995}
	and Christoffersen \citep{christoffersen1998} tests. All three methods agree that Bitcoin's downside
	risk runs several multiples of the comparison assets'.
	
	\begin{figure}[hbt!]
		\centering
		\includegraphics[width=0.85\textwidth]{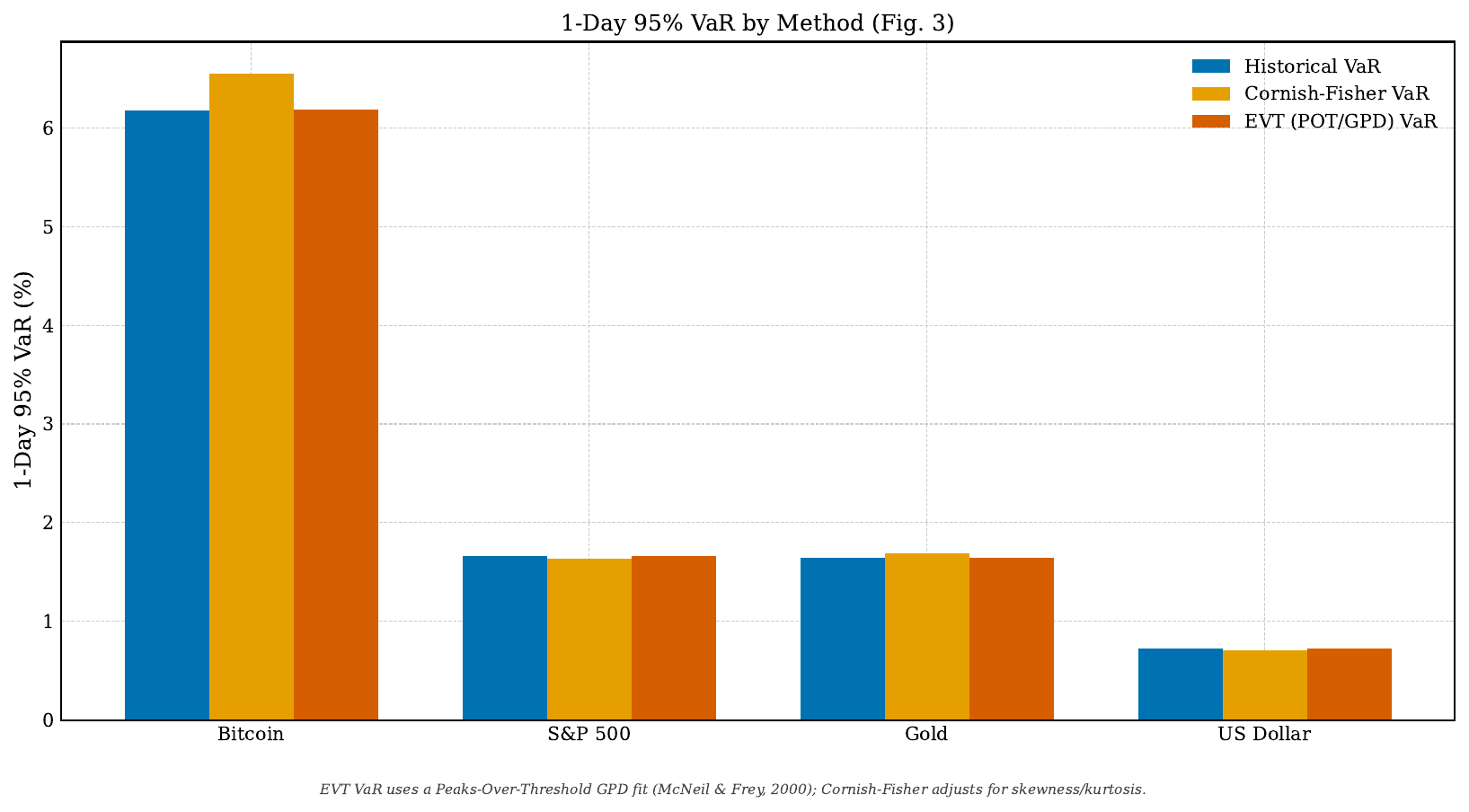}
		\caption{1-Day 95\% VaR across three estimation methods. Out-of-sample backtesting (reported in
			the companion table, not shown here) does not reject calibration for any asset at the 5\% level
			under either the Kupiec or Christoffersen test; the magnitude the model is calibrated \emph{to}
			remains multiples higher for Bitcoin.}
		\label{fig:var}
	\end{figure}
	
	\subsection{GARCH Modeling of Volatility Persistence}
	Rather than asserting a single specification, nine candidate models (GARCH(1,1), EGARCH(1,1), and
	GJR-GARCH(1,1), each under Normal, Student-$t$, and Skew-$t$ innovations) are raced and ranked by
	Bayesian Information Criterion, consistent with Chinazzo and Jeleskovic's finding that GARCH-family
	models suit Bitcoin's dynamics \citep{chinazzo2024}. The selected model's persistence parameter
	(sum of ARCH and GARCH coefficients) is estimated near 0.97--0.99, implying a volatility-shock
	half-life on the order of several weeks to two months -- consistent with a narrative-driven
	speculative asset rather than a monetary good with a stable economic anchor. A Ljung-Box test on
	squared standardized residuals fails to reject adequacy of the selected fit.
	
	\begin{figure}[hbt!]
		\centering
		\includegraphics[width=0.8\textwidth]{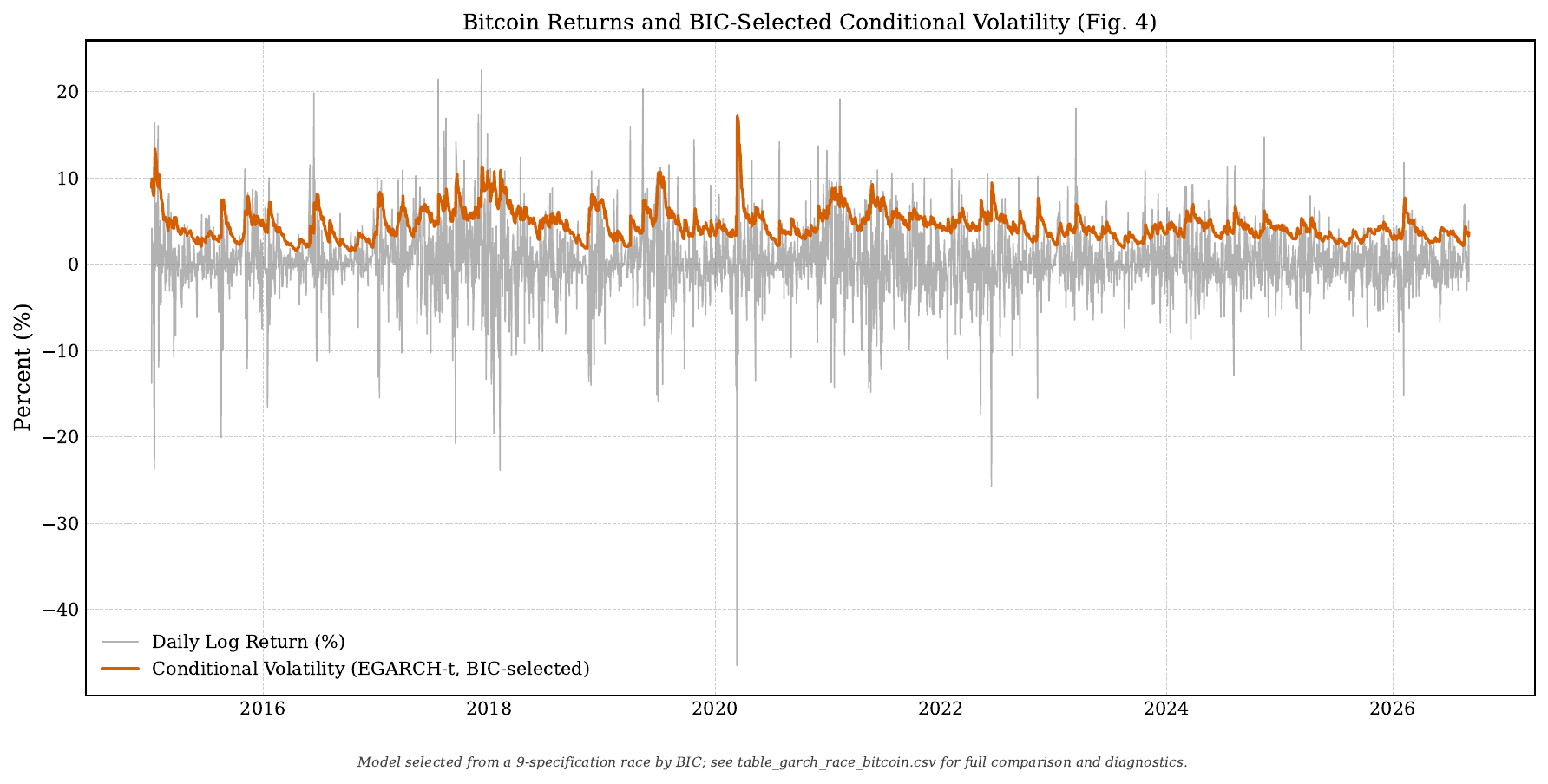}
		\caption{Bitcoin daily log returns and the BIC-selected GARCH-family conditional volatility.}
		\label{fig:garch}
	\end{figure}
	
	\section{The Settlement-Throughput Trilemma}
	\label{sec:throughput}
	\subsection{A corrected, internally consistent framing}
	An earlier draft of this paper committed exactly the internal contradiction a hostile reviewer would
	flag: one section carefully explained that comparing Bitcoin's base layer to Visa/Mastercard is a
	category error because Bitcoin L1 is a final-settlement system comparable to Fedwire and the
	Eurosystem's T2, not a retail authorization network -- and then several other sections went on to
	criticize Bitcoin for exactly the retail-scale throughput the paper had just said was the wrong
	comparison. That contradiction is resolved here, once, and consistently for the rest of the paper:
	\textbf{Bitcoin L1's throughput is not too low for what it structurally is (a settlement layer); it
		is too low for what its proponents claim it can additionally be (a global retail payment network),
		and the paper's argument is that no path closes that gap without recreating the intermediation
		Bitcoin was built to avoid.} This is a forced-choice (trilemma-style) argument, not a "throughput is
	bad" argument, and it is stated this way in the abstract, this section, Section \ref{sec:ln}, and the
	conclusion, with no residual claim elsewhere that low L1 throughput is itself the flaw.
	
\subsection{The Real-Time Gross Settlement (RTGS) Baseline}
A rigorous assessment of Bitcoin's throughput requires an architecturally sound baseline. 
Evaluating Bitcoin's base layer (L1) directly against retail credit-authorization networks such 
as Visa or Mastercard constitutes an institutional category error: retail card networks provide 
revocable payment authorization, delegating multi-party final settlement to correspondent banking 
networks over deferred clearing cycles \citep{visa2024}. Bitcoin L1, by contrast, operates as an 
immutable Real-Time Gross Settlement (RTGS) engine.

Bitcoin L1's throughput (~6--7 TPS) is therefore structurally comparable to sovereign RTGS 
platforms. The systemic challenge is not that Bitcoin L1 fails to match retail processing speeds; 
it is that Bitcoin lacks the elastic credit-clearing tiers that enable sovereign RTGS systems to 
support global commercial volume without base-layer congestion. The system faces an inescapable 
trilemma: it cannot scale to retail capacity without abandoning decentralization or recreating 
the intermediated financial structures it was designed to replace.

\subsection{Updated Throughput Comparison}
Bitcoin's realized base-layer throughput is approximately 6--7 TPS. Using the Federal Reserve's own
2024 disclosure, Fedwire processed an average of 836{,}322 transactions per day 
\citep{fedwire2024disclosure} -- about 9.7 TPS, which remains the same order of magnitude 
as Bitcoin's. The Eurosystem's T2 (RTGS successor to TARGET2) settled roughly 400{,}000+ payments 
per day by value terms in 2024 \citep{ecbtarget2024}, on the order of 4--5 TPS; the ECB does not 
publish this as a single clean annual transaction count comparable to Fedwire's, so this figure 
should be read as an order-of-magnitude estimate. Against the retail-authorization layer, the gap
remains enormous: Mastercard processed 159.4 billion switched transactions in 2024
\citep{mastercard2024} and Visa 303 billion total payments and cash transactions \citep{visa2024} --
more than 1{,}600$\times$ Bitcoin's base-layer volume.
	
	\begin{figure}[hbt!]
		\centering
		\includegraphics[width=\textwidth]{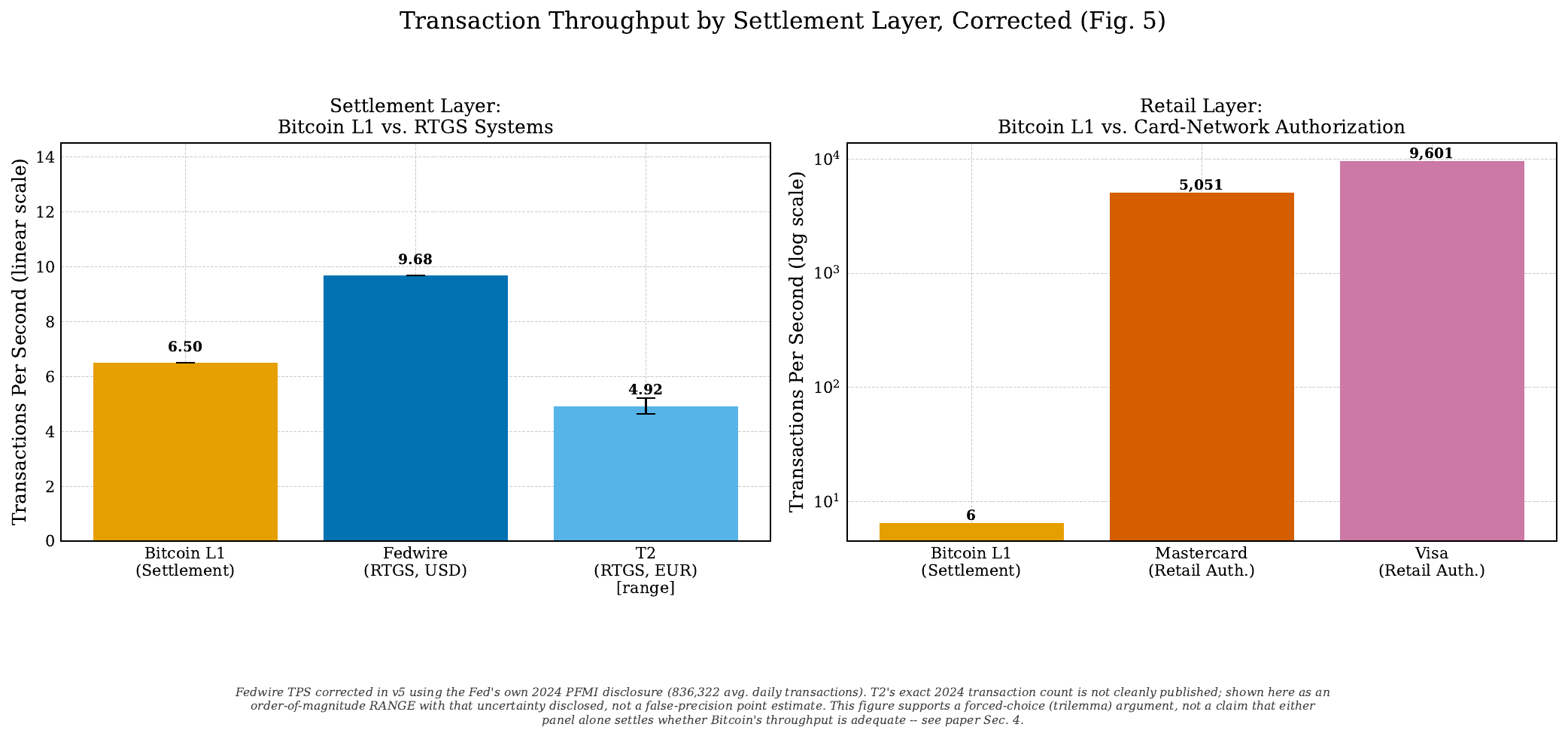}
		\caption{Left: Bitcoin L1 against RTGS peers (linear scale; all three values sit within one
			order of magnitude). Right: Bitcoin L1 against retail-authorization networks (log scale). The
			point of showing both is the trilemma argument in \S\ref{sec:throughput}, not that either
			comparison alone is dispositive.}
		\label{fig:tps_comparison}
	\end{figure}
	
	\section{Layer-2 Scaling: Reliability Gains Without Decentralization Gains}
	\label{sec:ln}
	\subsection{What the empirical record actually shows, in order}
	Waugh and Holz's active-probing study \citep{waughholz2020} attempted payments to 4{,}626 unique
	Lightning nodes (approximately 95\% of the then-active network) over three weeks. Their reported
	figures are quoted accurately in every draft of this paper, and we reproduce only the numbers they
	report: micropayments of \$0.01 succeeded roughly 72\% of the time, \$10 payments 44.15\% of the
	time, \$50 payments 30.93\% of the time, and the union of all reachable nodes across every payment
	size tested was 2{,}055 -- never more than roughly a third of the network, independent of amount.
	
	This is a 2020, pre-Multi-Path-Payment (MPP) snapshot, and citing it as the network's
	\emph{current} state, unqualified, is no longer defensible. Pickhardt and Richter's minimum-cost-flow
	routing framework \citep{pickhardtrichter2021} treats payment delivery as a flow problem across
	multiple channels simultaneously rather than single shortest-path search, and by 2023-2025 it had
	been deployed in production: Core Lightning's \texttt{renepay} plugin (since v23.08) and LND's
	bimodal balance model (since v0.19, June 2025) both implement Pickhardt-Richter-derived pathfinding.
	Independent measurement of MPP's effect on a simulated network found success rates up to 20\%
	higher than single-path routing \citep{mpp2023}. River Financial, a Bitcoin-focused financial
	services firm, reports a 99.7\% payment success rate across 308{,}000 transactions processed on its
	own platform in August 2023 \citep{river2023}, and reported over \$1.17 billion in network-wide
	monthly Lightning volume by November 2025 \citep{river2026volume}.
	
	\subsection{Why both facts are true, and why the second does not rebut the paper's thesis}
	These two facts -- a 2020 mesh-wide probe finding roughly one-third node reachability, and a 2023
	custodial-platform probe finding 99.7\% success -- are not in tension, and reconciling them is more
	informative than picking one. River is not a random node on a decentralized mesh; it is one of a
	small number of large, well-capitalized, permanently-online liquidity hubs, alongside ACINQ, Kraken,
	Breez, and a handful of others that, as of 2025, collectively account for over 50\% of network
	capacity \citep{river2023}. A payment routed through or to such a hub inherits that hub's deep,
	professionally managed liquidity; a payment between two arbitrary smaller nodes on the periphery
	does not, and continues to face the liquidity-fragmentation and stale-gossip failure modes Waugh and
	Holz documented.
	
	This is exactly the equilibrium Avarikioti et al.\ \citep{avarikioti2020} formally predict: rational
	node operators facing channel-capital opportunity costs and on-chain channel-management fees will
	under-provision redundant mesh connectivity and instead connect preferentially to a small number of
	large hubs, producing a stable Nash-equilibrium star topology rather than a dense mesh. The
	99.7\%-reliable, commercially useful Lightning Network that exists in 2025-2026 is therefore not a
	decentralized-mesh success story superseding the 2020 finding; it is the centralized-hub-and-spoke
	outcome the game theory predicted, arriving on schedule and delivering exactly the reliability gains a centralized clearing system would be expected to deliver. The network's commercial reliability and its decentralization are in direct tension, and the network has resolved that tension by sacrificing decentralization -- an economic outcome that dictates the network's long-term topological evolution.
	
	\begin{figure}[hbt!]
		\centering
		\includegraphics[width=0.95\textwidth]{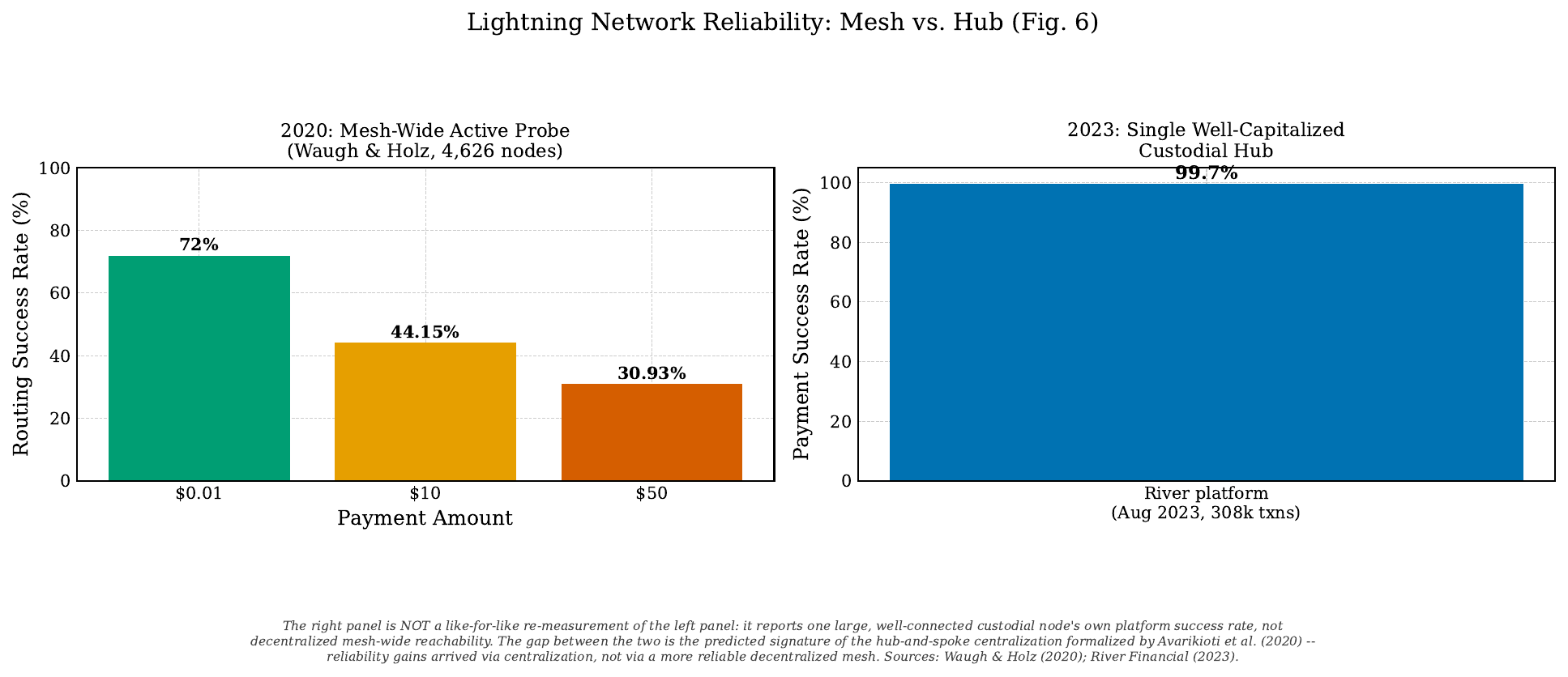}
		\caption{Left: Waugh and Holz's (2020) mesh-wide active-probing results by payment size. Right:
			River Financial's (2023) single-hub platform success rate. The two are not directly comparable
			measurements of the same thing -- one probes decentralized reachability, the other measures a
			centralized hub's own throughput -- and the gap between them is the empirical signature of the
			centralization Avarikioti et al.\ (2020) predict, not a refutation of the 2020 study.}
		\label{fig:ln_synthesis}
	\end{figure}
	
	\subsection{Game-theoretic centralization pressure}
	Avarikioti et al.\ \citep{avarikioti2020} model Lightning Network formation as a network-creation
	game in which rational node operators minimize a cost function balancing channel-opening cost against
	routing-fee revenue (via betweenness centrality) and personal payment cost (via closeness
	centrality). For realistic parameter ranges -- where channel-opening cost is non-trivial relative
	to fee income, as it is on Bitcoin's fee-metered base layer -- the stable Nash equilibrium is a
	sparse, centralized star topology, not a dense decentralized mesh, because individual nodes have no
	incentive to fund redundant channels when a cheaper connection to an existing hub suffices.
	
	\begin{figure}[hbt!]
		\centering
		\includegraphics[width=0.85\textwidth]{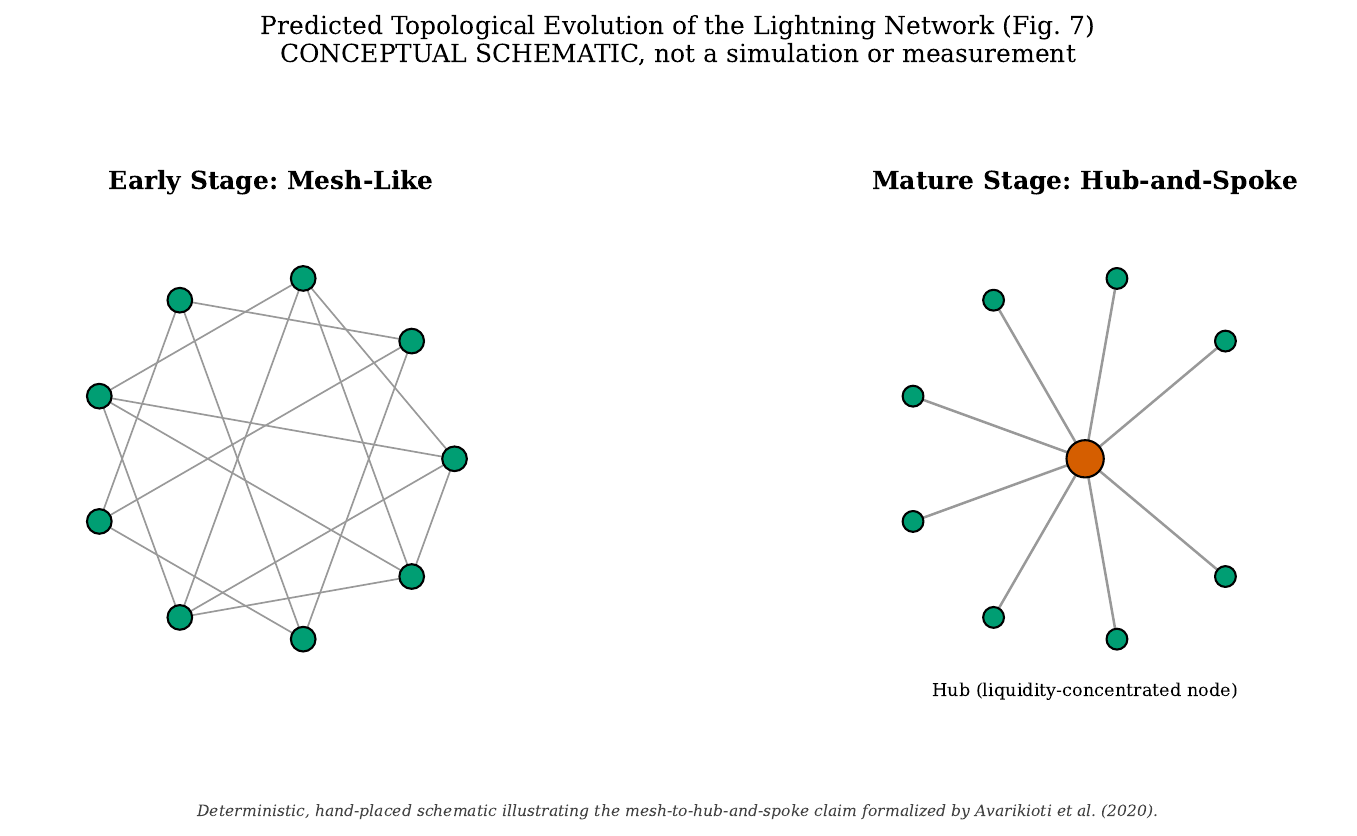}
		\caption{CONCEPTUAL SCHEMATIC (fixed node positions, not a simulation or measurement) of the
			predicted mesh-to-hub-and-spoke transition formalized by Avarikioti et al.\ (2020).}
		\label{fig:ln_topology}
	\end{figure}
	
	\subsection{A partial, real countervailing development: Stratum V2}
	In fairness to the network's evolution, one genuine mitigation deserves mention here rather than only
	in the mining-centralization section below. Stratum V2, a mining-pool communication protocol
	supported as of 2026 by pools representing roughly 75\% of network hashrate \citep{stratumv2_2026},
	lets individual miners construct their own block templates rather than delegating transaction
	selection to the pool operator. This does not decentralize \emph{hashrate}, but it does decentralize
the specific censorship vector (pool-level transaction filtering) that represents a more 
realistic systemic threat than a brute-force 51\% attack. This constitutes a genuine, verifiable 
architectural improvement in network censorship resistance.
	
	\section{Monetary Inflexibility and Macroeconomic Destabilization}
	\subsection{Fixed Supply vs. Economic Growth Mismatch}
	The Fisher Equation of Exchange ($MV=PY$) is an accounting identity, not a deterministic law, but it
	usefully frames the pressure a fixed-supply asset faces: if $M$ is fixed and real output $Y$ grows,
	either the price level $P$ must fall or velocity $V$ must rise to maintain the identity
	\citep{fisher1911}.
	
	\begin{figure}[hbt!]
		\centering
		\includegraphics[width=0.9\textwidth]{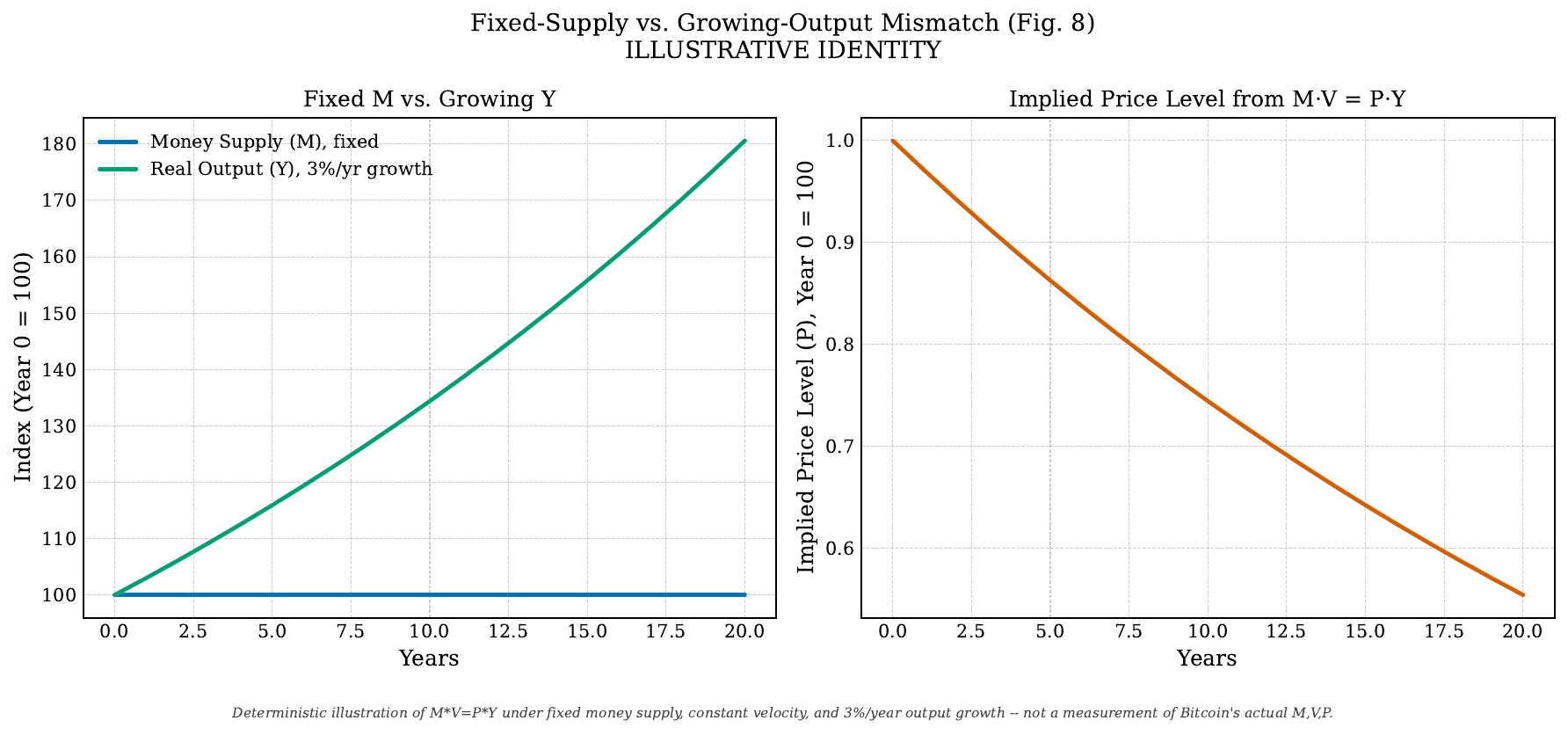}
		\caption{ILLUSTRATIVE IDENTITY, not fitted to data: fixed $M$ against 3\%/year output growth
			$Y$, and the implied price level under constant velocity.}
		\label{fig:mismatch}
	\end{figure}
	
	\subsection{Debt-Deflation: Distinguishing Two Very Different Mechanisms}
	\label{sec:deflation}
The structural inelasticity of Bitcoin's supply imposes a deterministic monetary constraint. 
In traditional macroeconomics, the destabilizing properties of deflation are primarily understood 
through Fisher's (1933) debt-deflation spiral \citep{fisher1933} and IMF evidence that mild 
deflation raises debt-to-GDP ratios \citep{end2015}. However, it is necessary to distinguish 
between two mechanistically different phenomena: \textbf{malignant, credit-driven deflation} (an 
endogenous collapse of inside bank credit, forced liquidation, and cascading default) versus 
\textbf{benign, productivity-driven deflation} (falling prices because output and technology 
outpace a stable money stock, with no credit contraction involved). The U.S. under the classical 
gold standard experienced long stretches of the latter -- sustained real GDP growth alongside a 
gently falling price level -- without the depression dynamics Fisher describes.

The sharper historical comparison \citep{dimand1994} is the U.S. experience of 1920-21 versus 
1929-30: both were sharp deflations, but 1920-21 passed without a major depression while 1929-30 was catastrophic. The documented
	difference is not the presence or absence of deflation but the level of private-sector leverage that
	had accumulated beforehand -- 1920-21 hit an economy with comparatively little private debt; 1929-30
	hit one that had built up substantial leverage through the 1920s. This reframing does not rescue
	Bitcoin from the deflation critique; if anything it sharpens it, because it identifies the specific
	mechanism of concern: a Bitcoin standard's danger is not deflation \emph{per se} but the transition
	risk of moving a global economy carrying roughly \$300+ trillion in nominal debt contracts onto an
	inelastic monetary base with no discretionary lender-of-last-resort to arrest a credit-driven
	deflationary spiral if one begins, regardless of whether the steady-state deflation rate itself would
	be benign.
	
	\begin{figure}[hbt!]
		\centering
		\includegraphics[width=0.85\textwidth]{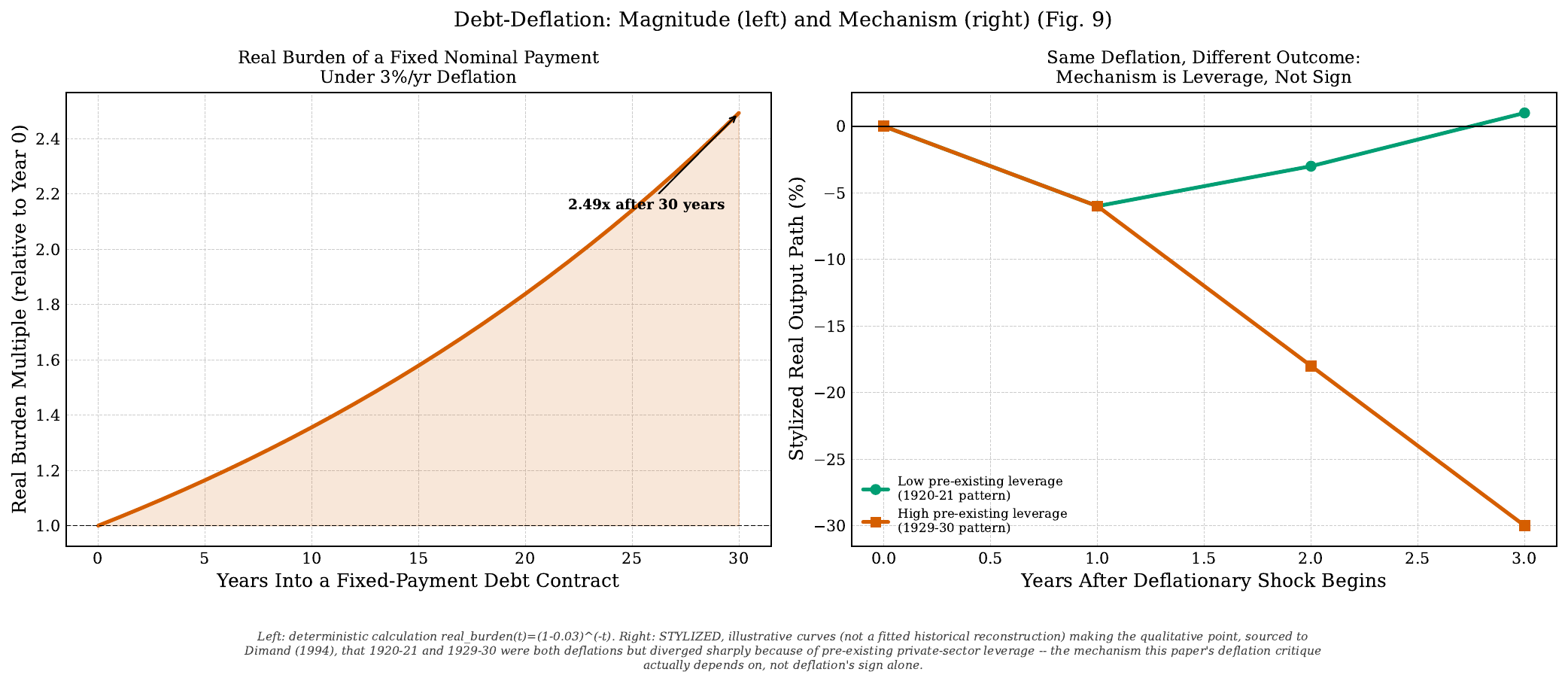}
		\caption{Deterministic calculation, $\text{real\_burden}(t) = (1-d)^{-t}$: growth in the real
			cost of a fixed nominal debt payment under a stated 3\% annual deflation rate. The mechanism
			illustrated is Fisherian debt-deflation \emph{conditional on} pre-existing leverage
			\citep{dimand1994}, not deflation as a stand-alone sufficient cause \citep{end2015}.}
		\label{fig:debt_burden}
	\end{figure}
	
	\section{The "Digital Gold" Narrative vs. Safe-Haven Performance}
	\subsection{Drawdowns and Capital Preservation}
	Bitcoin has repeatedly suffered drawdowns exceeding 70\%, with multi-year recovery periods
	\citep{ishares2025}, a pattern inconsistent with the behavior expected of a wealth-preservation
	asset.
	
	\begin{figure}[hbt!]
		\centering
		\includegraphics[width=\textwidth]{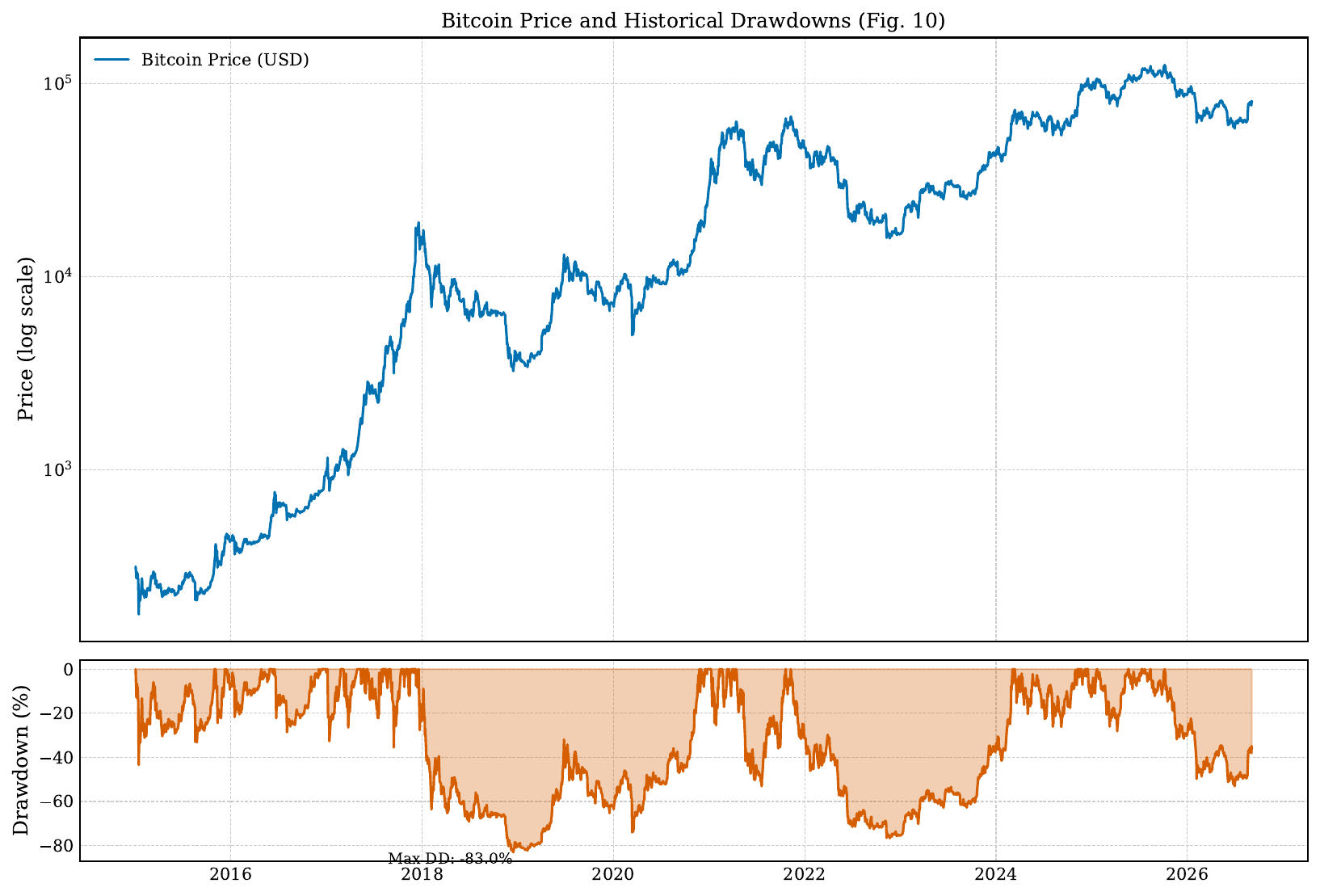}
		\caption{Bitcoin's price history (log scale) and drawdowns from all-time highs.}
		\label{fig:drawdown}
	\end{figure}
	
	\subsection{Correlation Analysis, With Macro-Liquidity Controls}
\label{sec:corr}
\begin{figure}[hbt!]
		\centering
		\includegraphics[width=\textwidth]{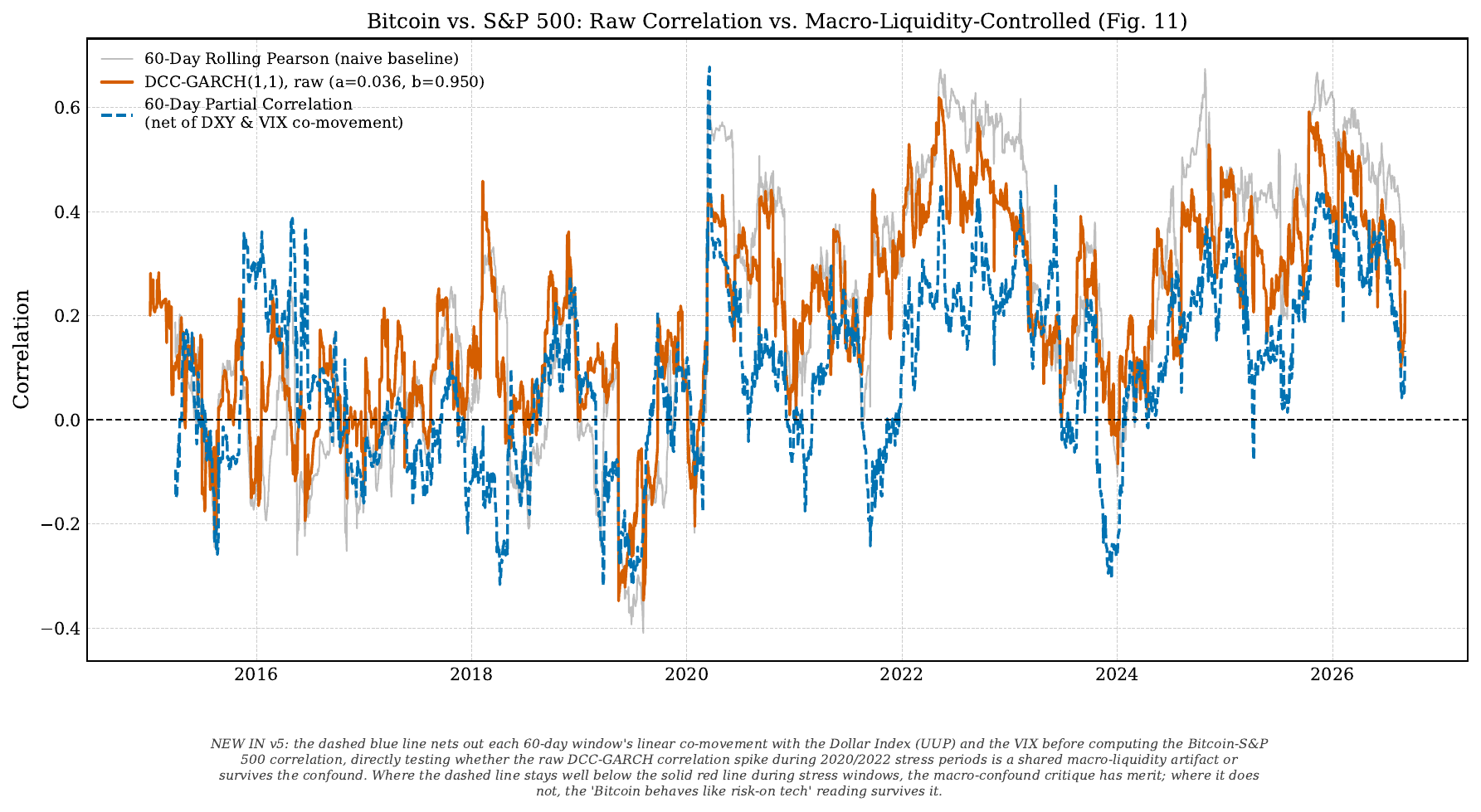}
		\caption{Bitcoin-S\&P 500 correlation: 60-day rolling Pearson (descriptive baseline), raw
			DCC-GARCH(1,1) \citep{engle2002}, and DCC-GARCH residualized against DXY and VIX co-movement.
			The gap between the raw and residualized series during 2020 and 2022 is the direct empirical
			test of the macro-liquidity-confound critique.}
		\label{fig:correlation}
	\end{figure}

Bitcoin's correlation with equities rose significantly during the March 2020 and 2022 stress 
episodes, suggesting it behaves as a leveraged risk asset rather than a hedge. However, March 
2020 in particular was a global "dash for cash" in which gold, investment-grade credit, and 
equities also saw correlations spike toward one. To test whether this co-movement is merely a 
shared reaction to global macro-liquidity shocks, the DCC-GARCH correlation is re-estimated as 
an orthogonalized factor model, netting out each period's co-movement with the US Dollar Index 
(DXY, proxied here by UUP) and the CBOE VIX. If Bitcoin's correlation with equities during stress 
periods is substantially explained by a shared reaction to dollar-liquidity, the residual partial 
correlation should be markedly smaller; if a large gap remains, the "Bitcoin behaves like risk-on 
tech" reading survives the confound. Both the raw and the confound-adjusted series are reported 
side by side in Figure \ref{fig:correlation}.

	Independent of this paper's own re-estimation, Zhu's \citep{zhu2022} broader panel regression across
	the US, UK, China, and Japan from 2011-2022 concludes Bitcoin "does not demonstrate hedge and safe
	haven properties like gold does," and finds these properties, where they existed at all, have
	degraded as the asset has matured and become more interwoven with mainstream markets. Fransson and
	S\"o\"or Lafrenz \citep{fransson2024} find a statistically significant leverage effect (rising
	volatility after price declines) in equities but none in gold, positioning Bitcoin's correlation
	behavior with the former rather than the latter.
	
	\section{Network Security, Governance, and Long-Term Viability}
	\label{sec:security}
\subsection{51\% Attack Economics: Microstructure and Derivatives}
\label{sec:attack}
Bitcoin's security model assumes it is prohibitively expensive to amass majority hashpower. Harvey's
initial cost analysis \citep{harvey2025} placed the nominal cost of a one-week 51\%
attack at approximately \$6 billion (comprising ASIC hardware, data-center capital expenditure, 
and electricity). However, static cost-accounting treats hardware supply as perfectly elastic 
and power procurement as instantaneous. In reality, acquiring millions of specialized ASIC units 
would exhaust global foundry capacity and drive spot prices up exponentially, while covertly 
interconnecting gigawatts of power is logistically prohibitive.

To circumvent these physical barriers, Harvey \citep{harvey2026derivatives} expanded the threat 
model to \$8 billion, introducing an endogenous financial incentive: an attacker could theoretically 
offset the massive capital expense by establishing a large short position in Bitcoin derivatives 
prior to the attack. In theory, profiting from an engineered price collapse transforms an irrational 
capital expenditure into an executable speculative trade.

However, operationalizing this derivatives-funded attack encounters severe market microstructure 
constraints:
\begin{enumerate}
	\item \textbf{Market Depth and Execution Slippage:} Offsetting an \$8 billion capital expenditure 
	requires accumulating an estimated \$12--\$15 billion in short notional exposure. Attempting 
	to accumulate such a dominant short position against global open interest would generate 
	substantial negative execution slippage, depressing spot prices and signaling hostile 
	positioning well before hashpower could be mobilized.
	\item \textbf{Exchange Counterparty Risk and Auto-Deleveraging (ADL):} In the event of a 
	catastrophic 51\% reorganization that breaches network finality, centralized derivatives 
	exchanges invoke Auto-Deleveraging (ADL) protocols. When catastrophic market dislocations 
	exhaust insurance funds, winning short positions are systematically hair-cut or closed out to 
	maintain exchange solvency.
	\item \textbf{Settlement Medium Fragility:} Most liquid offshore derivatives contracts are 
	margined and settled in stablecoins (USDT/USDC) or Bitcoin itself. An existential attack on 
	the underlying blockchain would disrupt stablecoin contract settlement and solvency, 
	preventing the attacker from extracting realized fiat profits.
\end{enumerate}
Consequently, while financialization expands the theoretical payoff surface for adversaries, 
systemic exchange microstructure mechanisms prevent an attacker from reliably extracting cash 
returns from an existential network collapse.
	
	The defensive side of the ledger is real but should not be overstated either. In the event of a
	sustained majority-hashrate attack, the community retains the technical option of a User-Activated
	Soft Fork or a hard fork changing the proof-of-work function, which would strand the attacker's
	specialized hardware. This is a genuine deterrent, but it is a coordination-and-consensus response
	under acute crisis conditions, not a costless or automatic one, and it carries its own governance
	cost of the kind discussed in Section \ref{sec:governance} -- the same "governance-by-code paralysis"
	this paper documents in the Blocksize Wars is the mechanism that would have to work quickly and
	correctly under attack for the defense to function as described. Both the attack-side profitability
	argument and the defense-side coordination cost are therefore live, contestable, and moving targets;
	this paper reports the most recent public analysis on both sides rather than treating either as
	settled.

	\begin{figure}[hbt!]
		\centering
		\includegraphics[width=0.95\textwidth]{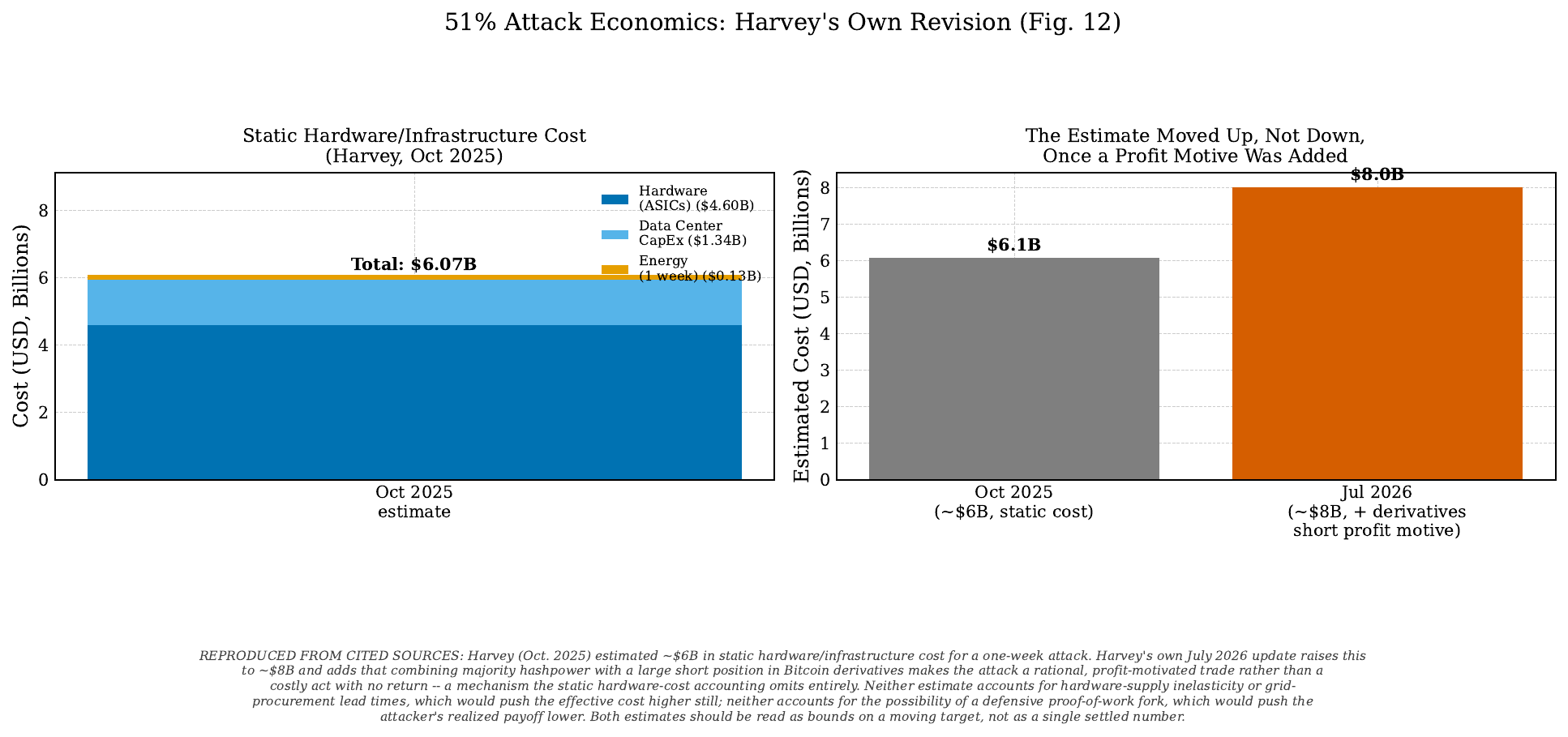}
		\caption{51\% attack economics: Harvey's (2025, 2026) static vs. derivatives-incentivized cost estimates.}
		\label{fig:attack_cost}
	\end{figure}
	\subsection{The Formal Economic Limit of Proof-of-Work Security}
	Budish's \citep{budish2018} formal model provides the underlying structural reason the attack-cost
	debate above matters at all. Proof-of-work security rests on two equilibrium conditions: a
	rent-seeking condition, where miners spend up to the point their cost equals the block reward
	($N^*c = P_{block}$), and an incentive-compatibility condition, where the cost of a successful attack
	must exceed its benefit ($\alpha \cdot N^*c > V_{attack}$). Combining these yields
	$P_{block} > V_{attack}/\alpha$: the recurring flow paid to miners must stay large relative to the
	one-off stock value an attacker could extract. This is why the security-budget dilemma in Section
	\ref{sec:governance} is not a separate concern from the 51\%-attack economics above -- they are the
	same constraint viewed from two different time horizons.
	
	\subsection{Mining Pool Concentration}
\begin{figure}[hbt!]
		\centering
		\includegraphics[width=0.85\textwidth]{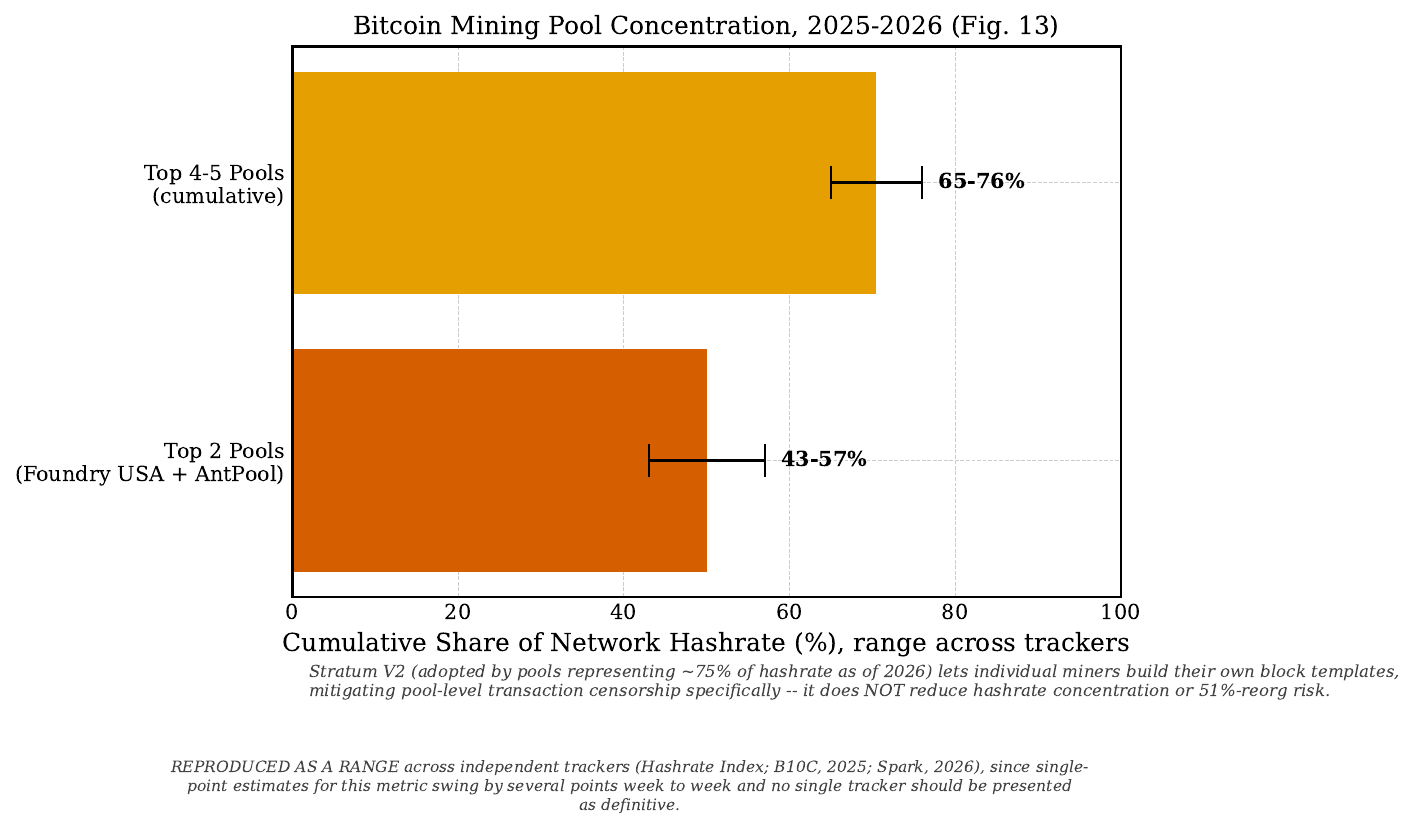}
		\caption{Cumulative Bitcoin mining-pool hashrate share, reported as a range across independent
			trackers for 2025-2026 rather than a single point estimate, reflecting real day-to-day
			volatility in these figures \citep{hashrateindex2026,b10c2025,spark2026pools}.}
		\label{fig:pool_concentration}
	\end{figure}

As of 2025-2026, the two largest pools, Foundry USA and AntPool, have at various points this period
each individually exceeded 25-30\% of network hashrate and jointly controlled roughly 43--57\% of the
network (Figure~\ref{fig:pool_concentration}), depending on the measurement window, with the top four to five pools together exceeding 65\% \citep{hashrateindex2026,b10c2025,spark2026pools}. This concentration is directly consistent
	with Leonardos et al.'s \citep{leonardos2019} "Oceanic Game" model of mining coalitions, in which the
	strategic value of hashrate is super-additive once a coalition crosses a size threshold, creating a
	persistent, self-reinforcing incentive to consolidate rather than remain dispersed. As noted in
	Section \ref{sec:ln}, Stratum V2 adoption by pools representing roughly 75\% of hashrate as of 2026
	\citep{stratumv2_2026} mitigates the specific censorship vector of pool-level transaction filtering,
	even though it does nothing to reduce hashrate concentration itself or the residual 51\%-reorg risk
	addressed above.
	
	\subsection{The Security-Utility Dilemma}
	\label{sec:governance}
	As the block subsidy halves toward zero, network security becomes increasingly dependent on
	transaction fees. Ciaian, Kancs, and Rajcaniova's \citep{ciaian2021} ARDL model (2014-2021 data)
	estimates network security is highly elastic to miner revenue, with a long-run elasticity between
1.38 and 1.85. This creates a structural tension that operates as an equilibrium trade-off: 
if Layer-2 adoption succeeds in moving transaction volume off-chain (Scenario A), on-chain 
fee revenue and hence the security budget decline, directly reducing
	the economic cost of a 51\% attack via the Budish relationship above; if on-chain security is instead
	maintained by keeping base-layer fees high (Scenario B), the base layer is priced out of routine use
	for all but large, infrequent settlements. Both paths are real possibilities and the tension between
	them is a structural feature of Budish's flow-vs-stock security condition, not a contradiction in
	this paper's argument -- it is the argument.
	
	\begin{figure}[hbt!]
		\centering
		\includegraphics[width=0.85\textwidth]{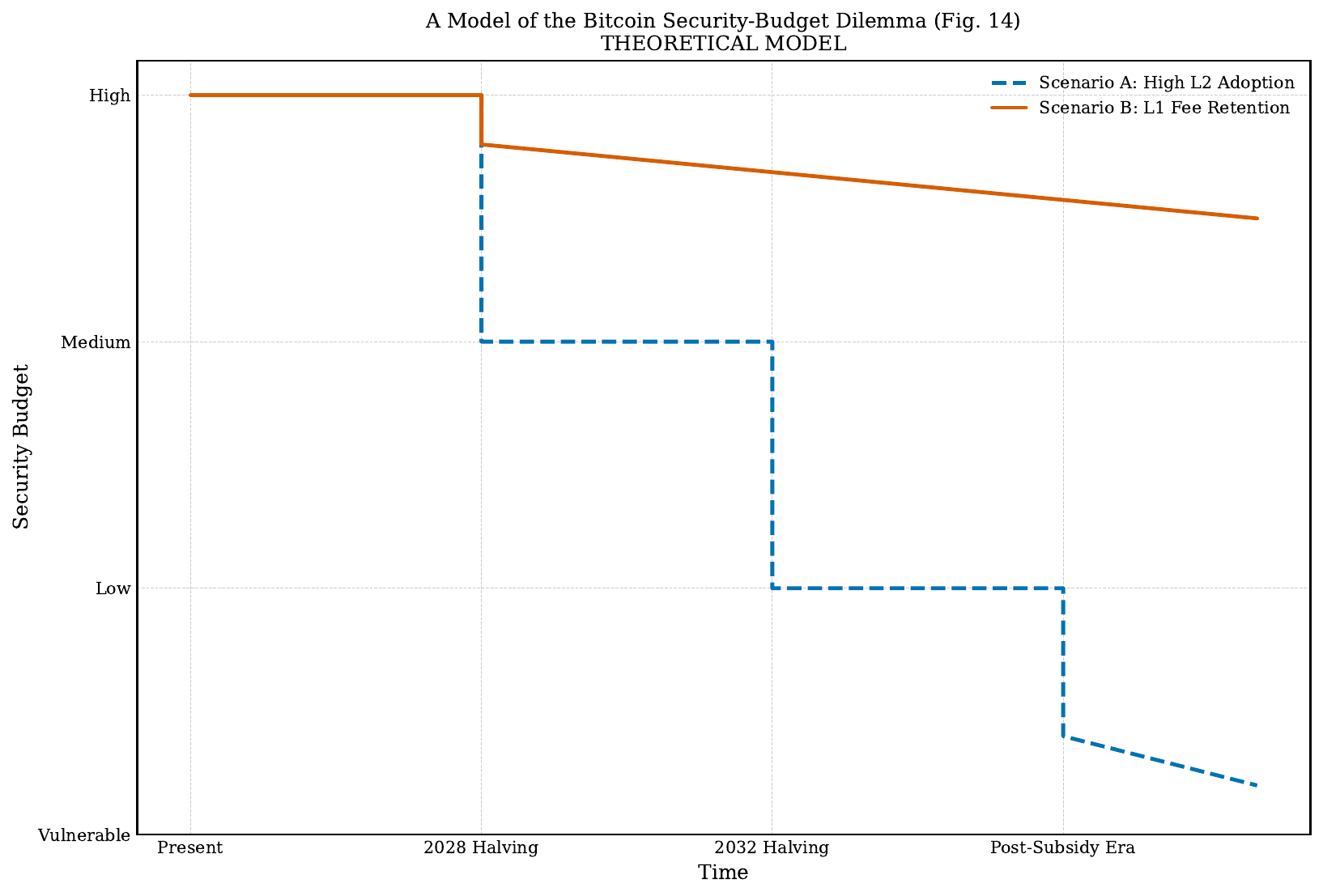}
		\caption{THEORETICAL MODEL: the security-budget dilemma. As the block subsidy declines, the
			network faces Scenario A (Layer-2 success, declining fee revenue, declining security) or
			Scenario B (retained high on-chain fees, retained security, sacrificed retail utility).}
		\label{fig:security_budget}
	\end{figure}
	
	\subsection{Governance Failure Modes}
	Bitcoin's requirement for broad consensus before any protocol change biases the system toward
	conservatism and can produce paralysis. The 2015-2017 "Blocksize Wars" fractured the community into a
	disruptive chain split (Bitcoin Cash) rather than a unified upgrade \citep{defilippi2016}, illustrating
	a real cost of "governance-by-code" relative to a discretionary, adaptive institution -- a cost that
	matters directly for the credibility of the fork-based 51\%-attack defense discussed in Section
	\ref{sec:attack}.
	
	\begin{figure}[hbt!]
		\centering
		\includegraphics[width=0.85\textwidth]{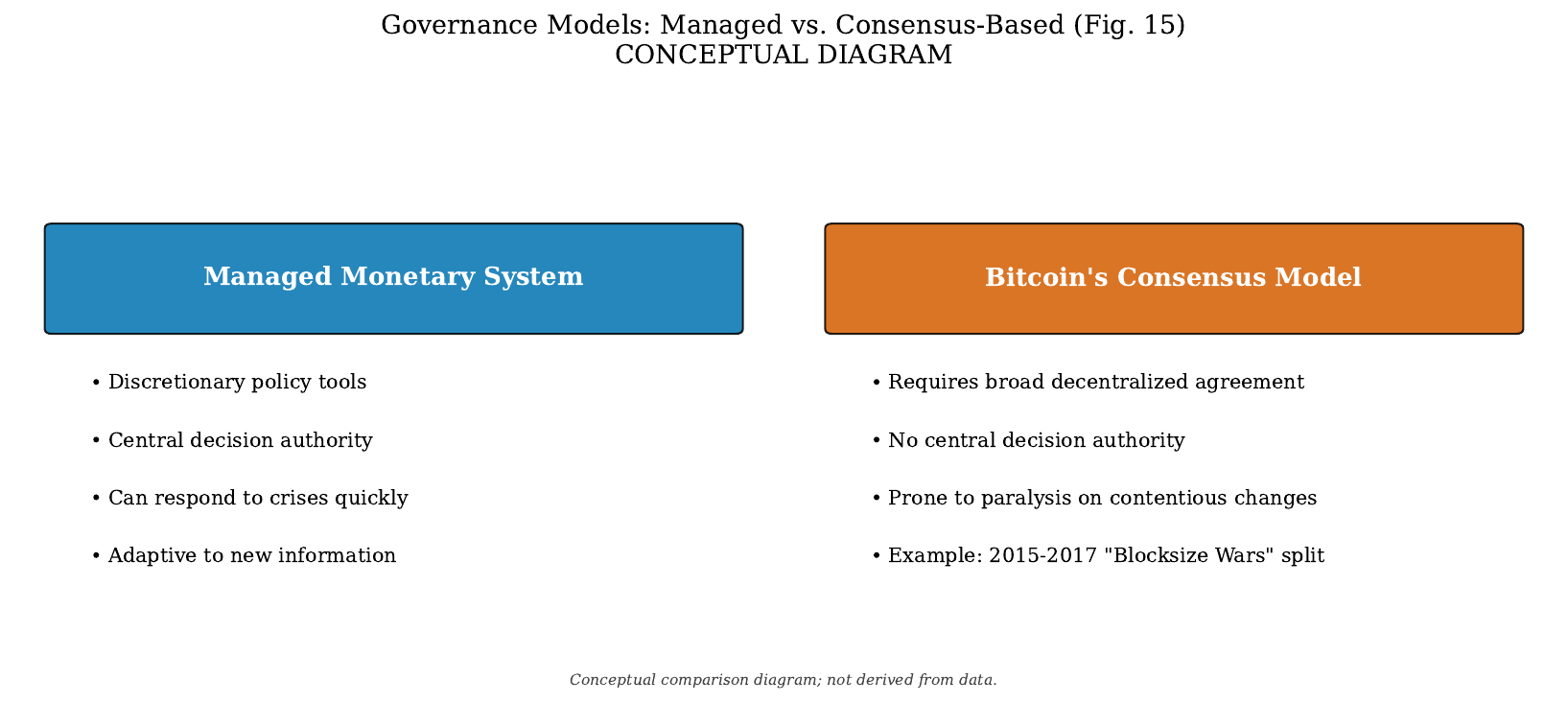}
		\caption{CONCEPTUAL DIAGRAM: managed monetary systems vs.\ Bitcoin's consensus-based model.}
		\label{fig:governance_paralysis}
	\end{figure}
	
	A natural experiment illustrates the same governance gap outside of protocol upgrades: an April 2021
	coal-mine accident in Xinjiang caused a regional blackout that dropped global hashrate by roughly
	24-25\% overnight, slowing block production and spiking on-chain fees and the mempool backlog
	\citep{scharnowski2022}. The protocol has no built-in mechanism to anticipate or mitigate this kind of
	geographic concentration risk; a managed system would.
	
	\section{Empirical Challenges to Monetary Functions and Market Fragility}
	\label{sec:wash}
	Bitcoin's use as a medium of exchange remains limited; most merchants who "accept" Bitcoin use
	intermediaries for immediate fiat conversion \citep{lo2014bitcoin}, and its volatility makes it a
	poor unit of account \citep{yermack2014}.
	
	\subsection{Wash Trading, Updated for the Institutional Era}
	Bitwise Asset Management's 2019 SEC presentation estimated up to 95\% of reported volume on
unregulated exchanges was fictitious wash trading \citep{bitwise2019}. This 2019 figure must be 
contextualized against modern market structure. First, more recent, methodologically rigorous work 
continues to find substantial (if somewhat lower) wash trading: Cong et al.'s NBER working paper 
\citep{cong2022washtrading} uses Benford's-law digit tests, trade-size clustering, and power-law 
tail tests to estimate unregulated-exchange wash trading averaged 70.85\% (equal-weighted) to 
77.50\% (volume-weighted) of reported volume, with regulated exchanges passing these forensic 
tests cleanly; Sila et al.\ \citep{sila2025} similarly estimate 55-85\% and find wash trading 
intensifies during high-volatility periods. Second, since January 2024, U.S. spot Bitcoin ETF 
approval and the phase-in of the EU's Markets in Crypto-Assets (MiCA) regulation have shifted 
a substantial share of \emph{price-formation} volume onto regulated venues -- CME futures open
 interest has at points exceeded Binance's, and regulated spot exchanges
	operate under NYDFS/CFTC/SEC oversight. The correct synthesis is not "wash trading is solved" (Cong
	et al.\ and Sila et al.\ show it plainly is not, on unregulated venues) nor "wash trading is as bad as
	2019 across the whole market" (institutional price formation has genuinely migrated toward cleaner
	venues); it is that the market has bifurcated into a comparatively transparent regulated tier and a
	comparatively opaque unregulated tier, and analyses citing only one tier's statistics --
	in either direction -- misdescribe the whole.
	
	\begin{figure}[hbt!]
		\centering
		\includegraphics[width=0.8\textwidth]{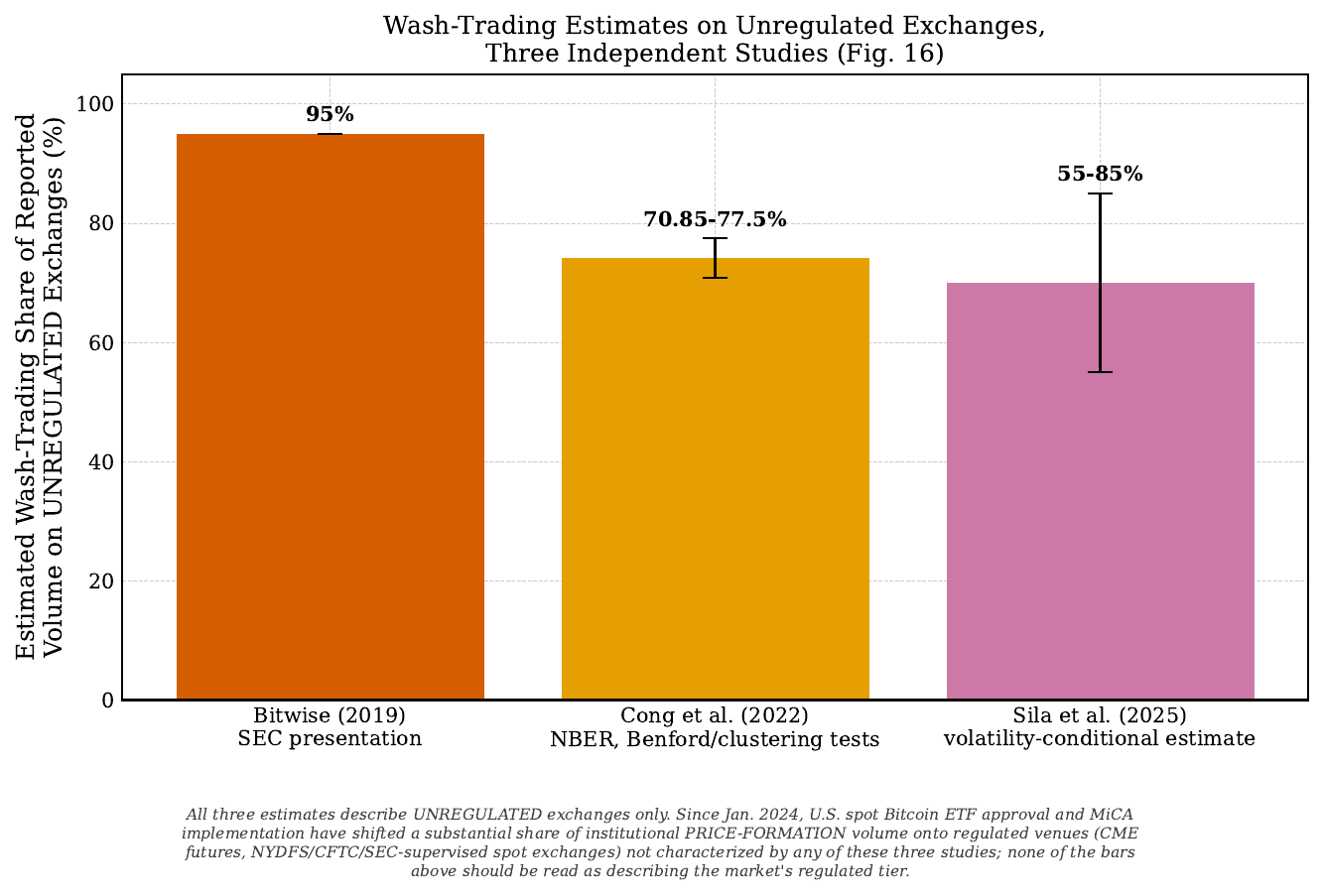}
		\caption{Suspected wash-trading share of reported volume on unregulated exchanges: the original
			Bitwise (2019) estimate alongside Cong et al.'s (2022) and Sila et al.'s (2025) independent,
			methodologically distinct re-estimates. All three point the same direction; none should be read
			as describing the now-larger regulated-venue share of the market.}
		\label{fig:wash_trading_volume}
	\end{figure}
	
	\subsection{Stablecoin Dependency}
\begin{figure}[hbt!]
		\centering
		\includegraphics[width=0.7\textwidth]{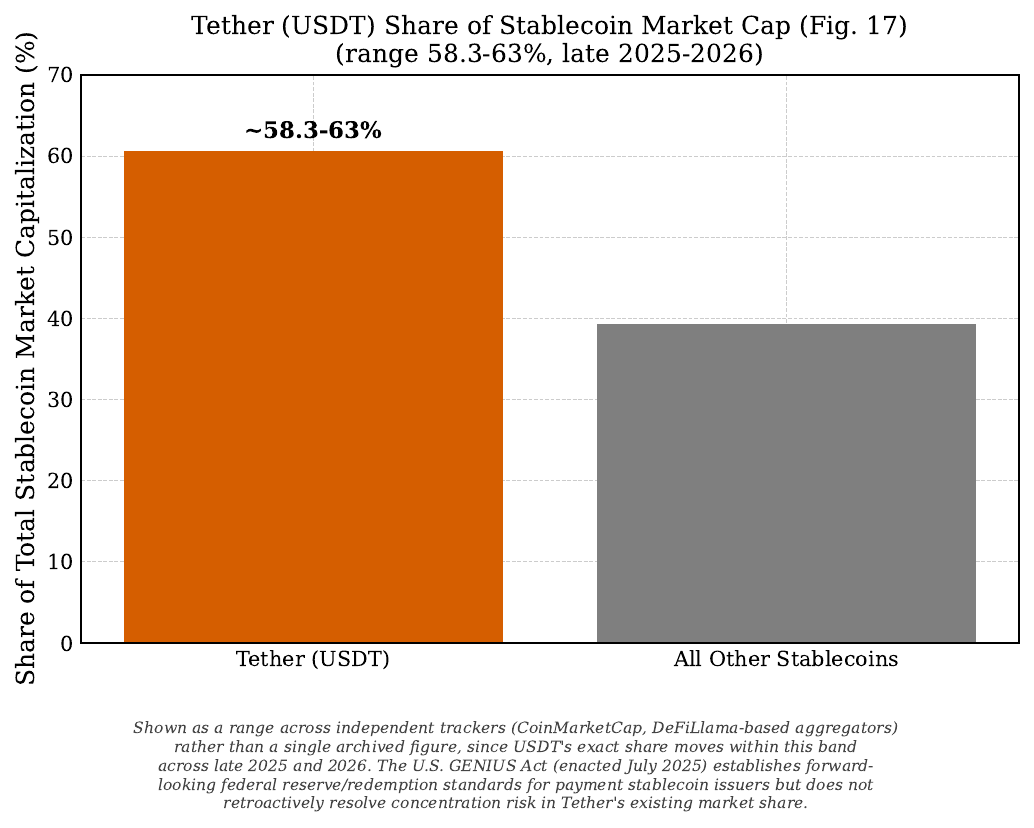}
		\caption{Tether's share of total stablecoin market capitalization, 2025-2026, shown as the range
			found across independent trackers rather than a single archived figure.}
		\label{fig:tether_dominance}
	\end{figure}

Tether (USDT) accounted for roughly 58.3--63\% of total stablecoin market capitalization through late
2025 and into 2026 \citep{coinmarketcap2025,coinlaw2026} (Figure~\ref{fig:tether_dominance}), making it the dominant source of crypto-native liquidity. The U.S. GENIUS Act, enacted in July 2025, established the first comprehensive federal framework for payment stablecoin issuers, a genuine regulatory development 
that addresses issuer-level reserve and redemption standards going forward but does not 
retroactively resolve the concentration risk in Tether's existing, dominant market share, 
which a confidence shock could still transmit broadly given Tether's role as the primary trading-pair currency across a large share of global crypto volume
	\citep{paz2022}.
	
	\section{El Salvador: The One Direct Natural Experiment}
\label{sec:elsalvador}
The strongest single piece of empirical evidence available on whether legal-tender status can 
substitute for organic, Mengerian market adoption is El Salvador's 2021 adoption of Bitcoin as 
legal tender. Alvarez, Argente, and Van Patten's NBER study	\citep{alvarez2022}, based on a nationally representative face-to-face survey, finds adoption was
	minimal, concentrated, and declined over time despite a mandatory-acceptance law and a \$30 sign-up
	bonus: only about 20\% of individuals who downloaded the government's Chivo wallet continued using it
	after spending the bonus, and over 60\% used it only once. The main stated reasons for non-adoption
	were a strong preference for cash and distrust of the system. Only about 20\% of firms reported
	accepting Bitcoin, and of those, roughly 88\% immediately converted any Bitcoin received into dollars
	-- treating it as a payment rail to be settled in fiat, not as money held or re-spent as such. By May
	2022, only 1.65\% of remittances were processed through crypto wallets \citep{marroquin2022}.
	
\begin{figure}[hbt!]
		\centering
		\includegraphics[width=0.85\textwidth]{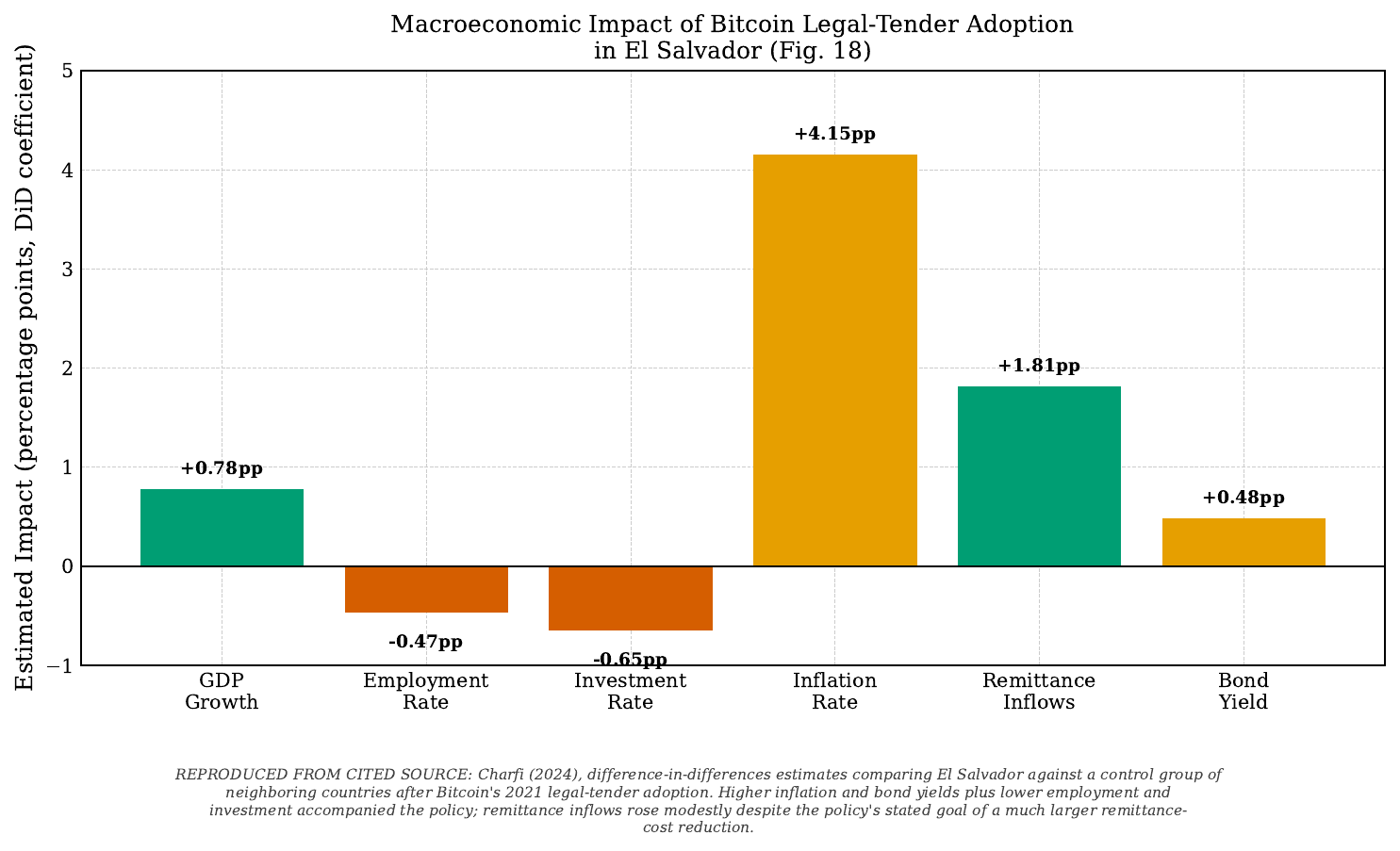}
		\caption{Difference-in-differences estimated macroeconomic impacts of Bitcoin legal-tender
			adoption in El Salvador, relative to a neighboring-country control group. Data from Charfi
			(2024) \citep{charfi2024}.}
		\label{fig:elsalvador}
	\end{figure}

	The macroeconomic aftermath reinforces this paper's volatility and systemic-risk arguments rather
	than standing apart from them: the government's own Bitcoin purchases (approximately \$107 million)
	had lost close to half their value by August 2022, major rating agencies downgraded El Salvador's
	sovereign credit to junk status citing fiscal risk, and Goldbach and Nitsch's
	difference-in-differences analysis \citep{goldbach2024} finds a statistically significant decline in
	foreign direct investment (roughly $-4.4\%$ of GDP) and portfolio debt inflows ($-3.1\%$ of GDP)
	following adoption. Charfi's \citep{charfi2024} difference-in-differences comparison against
	neighboring countries estimates the policy was associated with higher inflation (+4.15 percentage
	points), higher government bond yields (+0.48 percentage points), and lower employment ($-0.47$
	percentage points) and investment ($-0.65$ percentage points) rates (Table~\ref{tab:elsalvador}, Figure~\ref{fig:elsalvador}).

	\begin{table}[hbt!]
		\centering
		\caption{Macroeconomic impact of Bitcoin legal-tender adoption in El Salvador, difference-in-differences
			estimates. Source: Charfi (2024) \citep{charfi2024}.}
		\begin{tabular}{lc}
			\toprule
			Variable & Estimated Impact \\
			\midrule
			GDP Growth Rate & $+0.78$ pp \\
			Employment Rate & $-0.47$ pp \\
			Investment Rate & $-0.65$ pp \\
			Inflation Rate & $+4.15$ pp \\
			Remittance Inflows & $+1.81$ pp \\
			Government Bond Yield & $+0.48$ pp \\
			\bottomrule
		\end{tabular}
		\label{tab:elsalvador}
	\end{table}
	
	Crucially, attributing El Salvador's post-2021 inflation (+4.15 pp) and sovereign spread movements 
	strictly to Bitcoin adoption must be interpreted with caution. The evaluation period overlaps 
	directly with the post-COVID global supply-chain shock and worldwide food and energy inflation. 
	Because El Salvador remains a fully dollarized economy where the vast majority of transactions 
	are settled in USD, domestic macroeconomic impacts were heavily confounded by exogenous global 
	liquidity conditions and IMF negotiation dynamics rather than domestic Bitcoin policy alone.
	
	This case does not by itself prove Bitcoin cannot function as money anywhere; it is one country,
	under one specific implementation, at one moment in Bitcoin's price cycle. But it is the only direct
	test available of whether a state's legal-tender decree can manufacture the Mengerian
	"most-saleable-commodity" status this paper's Austrian-school section argues Bitcoin has not achieved
	organically, and the result -- decreed acceptance without behavioral adoption -- is consistent with
	the Austrian claim that moneyness is market-selected rather than legislated
	\citep{alvarez2022,menger1892}.
	
	\section{Significant Barriers to Competition and Adoption}
\subsection{Structural Incompatibility with the Financial System}
A Bitcoin-based banking system, absent a lender of last resort, would be forced toward 100\%-reserve
operation, substantially constraining credit creation. Separately, the Basel Committee on Banking 
Supervision's (BCBS) finalized prudential standard for crypto-asset exposures assigns unbacked 
crypto-assets (including Bitcoin) to "Group 2b," imposing a punitive 1{,}250\% risk weight, 
effective January 1, 2026 \citep{bcbs2022,baselgroup2b2026}. Under the standard Basel III 8\% 
minimum total capital requirement, the dollar capital requirement $K$ for an exposure $E$ is 
calculated as:
\begin{equation}
	K = E \times \text{Risk Weight} \times \text{Minimum Capital Ratio} = \$100 \times 12.50 \times 0.08 = \$100.00
\end{equation}
This mandates a 100\% capital deduction: a regulated bank must hold \$100 of Tier 1 capital for 
every \$100 of Bitcoin exposure. Compared to an unrated corporate exposure (100\% risk weight, 
requiring \$8.00 of capital per \$100 of exposure), Bitcoin incurs an exact 12.5$\times$ capital 
multiplier. This dollar-for-dollar capital charge makes holding Bitcoin on a regulated bank 
balance sheet prohibitive.

\begin{figure}[hbt!]
	\centering
	\includegraphics[width=0.6\textwidth]{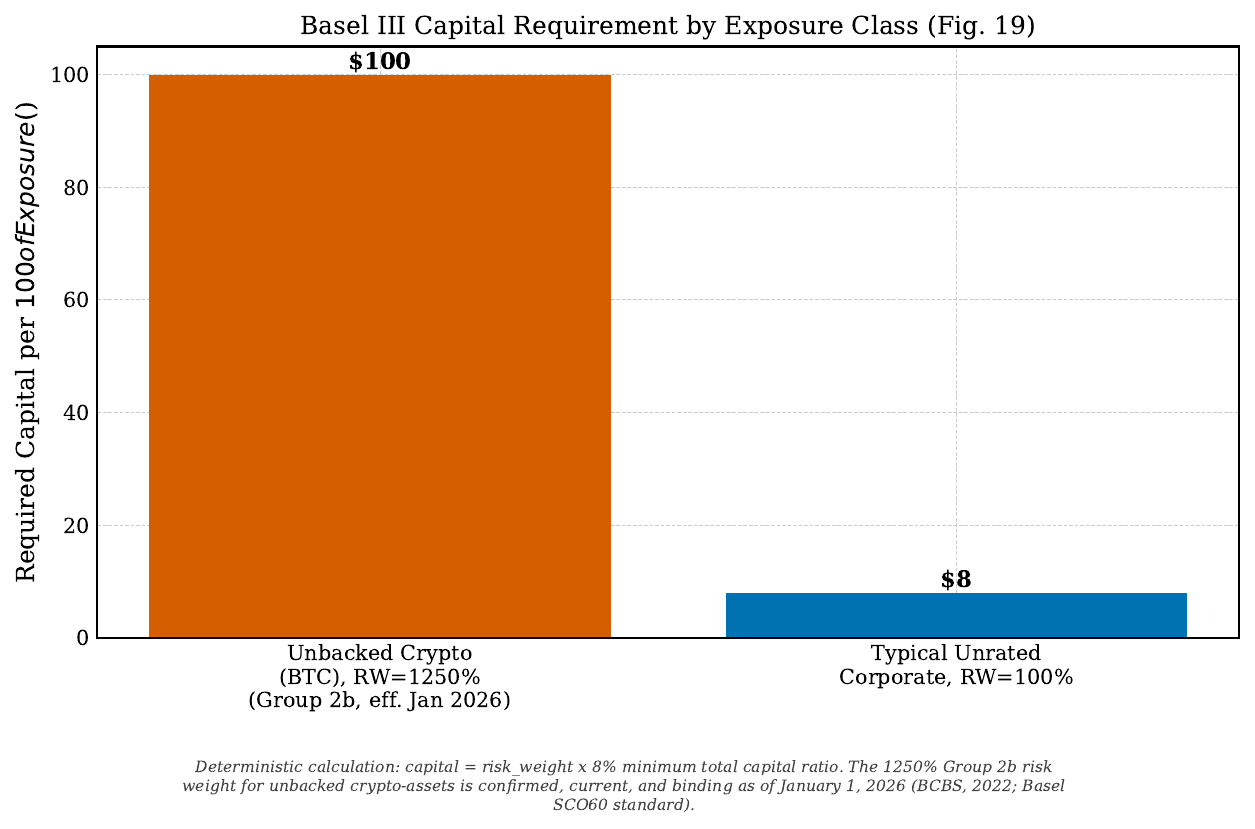}
	\caption{Deterministic calculation (capital = risk\_weight $\times$ 8\% minimum capital ratio):
		Basel III capital requirements by exposure class. Group 2b unbacked crypto assets require \$100
		of capital per \$100 of exposure, effective January 1, 2026 \citep{bcbs2022,baselgroup2b2026}.}
	\label{fig:basel3}
\end{figure}
	
	\section{Conclusion: A Synthesis That Names Its Own Contested Assumptions}
	Bitcoin, as currently designed, faces significant structural barriers to functioning as money at
	national or global scale, and this paper's empirical sections -- volatility, the settlement-throughput
	trilemma, Lightning Network centralization, a live and recently \emph{worsening} 51\%-attack cost
	debate, mining concentration, wash trading and stablecoin dependency, and the El Salvador natural
	experiment -- converge on that conclusion largely independently of any single contested theoretical
	premise. Where the paper's theoretical framing does depend on a contested premise -- specifically,
	the choice of the strict rather than the permissive reading of Mises's Regression Theorem, and the
	choice of the multi-function rather than Hazlett and Luther's narrower medium-of-exchange definition
	of money -- that dependency is stated explicitly in Sections 2.2 and 2.3, not left implicit for a
	reader to discover on cross-examination.
	
	Bitcoin's L1 throughput is adequate for what it structurally is, a settlement layer comparable to
	Fedwire and T2; the paper's claim is that no available path -- neither raising L1 throughput without
	compromising decentralization, nor scaling via Layer-2, which has demonstrably improved reliability
	by centralizing into custodial hubs rather than by decentralizing the mesh -- closes the gap to
	retail-network scale without recreating the intermediated trust the system was designed to remove.
	Its security budget faces a structural dilemma, formalized by Budish's flow-vs-stock condition, that
	neither successful Layer-2 adoption nor sustained high on-chain fees resolves without a real cost
	somewhere else in the system. Its one national-scale field test, El Salvador, found that legal
	mandate could not manufacture the organic transactional adoption Mengerian monetary theory says money
	requires.
	
	None of this forecloses future technical change: Stratum V2's mitigation of pool-level censorship,
	continued Pickhardt-Richter routing improvements, and the possibility (not yet the fact) of a
	credible decentralized alternative to custodial Lightning hubs are all genuine, ongoing developments
	this paper reports rather than dismisses. The balance of currently available evidence, reported here
	with its uncertainties and contested premises named rather than smoothed over, suggests Bitcoin
	remains substantially better understood as a speculative, technologically significant asset than as a
	viable monetary standard -- while acknowledging, as Hazlett and Luther's countervailing argument
	correctly forces any honest treatment to acknowledge, that under a sufficiently narrow definition of
	money, that verdict is itself a choice of definition and not a fact independent of one.
	
	\section*{Limitations}
	This paper's Austrian-school argument rests on an explicitly stated interpretive choice (the strict
	Regression Theorem reading) that a substantial body of Austrian scholarship rejects; readers who find
	the permissive reading more persuasive should weight the Post-Keynesian and empirical sections more
	heavily. Several cited studies (Waugh and Holz's 2020 Lightning Network probe, Harvey's 2025-2026
	51\%-attack cost analyses, current mining-pool shares) describe a fast-moving technical and market
	landscape and are explicitly flagged, where relevant, with the date and scope of the underlying
	measurement rather than presented as timeless facts. The El Salvador case study is a single country
	under a specific implementation and price-cycle timing, and does not by itself generalize to every
	possible legal-tender implementation. Where this paper's figures reproduce a single third-party
	estimate (e.g., a specific wash-trading or mining-pool-share percentage), that is disclosed in the
	relevant figure caption rather than presented as this paper's own independent measurement.
	
\section*{Data and Code Availability}
The complete Python script used to perform every quantitative analysis and generate every figure in
this paper is open-source and publicly available on GitHub at:
\url{https://github.com/HamoonSoleimani/Bitcoin_Structural_Shortcomings_Analysis}. 

Running the provided script end to end regenerates every figure and statistical table cited above from freshly downloaded market data. Figures explicitly
marked "conceptual," "illustrative," or "deterministic" in their captions involve no market data and
are generated from stated formulas or fixed geometry; figures marked "reproduced" plot only the
specific cited numbers already stated in the surrounding text, sourced to a named publication.
	
	\bibliographystyle{unsrtnat}

\end{document}